\documentclass[aps,superscriptaddress,prb,twocolumn]{revtex4-1}
\usepackage{amssymb,amsmath,mathptmx}
\usepackage{graphicx}
\usepackage{dcolumn}
\usepackage{bm}
\usepackage{hyperref}
\usepackage{color}
\usepackage[utf8x]{inputenc}
\usepackage{array}
\usepackage{ulem}
\usepackage{soul}

\newcolumntype{C}{>{\centering\arraybackslash}p{1em}}

\newcommand{\be}{\begin{equation}}
\newcommand{\ee}{\end{equation}}
\newcommand{\bea}{\begin{eqnarray}}
\newcommand{\eea}{\end{eqnarray}}

\begin{document}

\title{Revisiting the adiabatic limit in ballistic multiterminal
 Josephson junctions}

\author{R\'egis M\'elin}
\email{regis.melin@neel.cnrs.fr}

\affiliation{Universit\'e Grenoble-Alpes, CNRS, Grenoble INP, Institut
 NEEL, Grenoble, France}

\author{Asmaul Smitha Rashid}

\affiliation{Department of Electrical Engineering, The Pennsylvania
 State University, University Park, Pennsylvania 16802, USA}

\author{Romain Danneau}

\affiliation{Institute for Quantum Materials and Technologies,
 Karlsruhe Institute of Technology, Karlsruhe D-76021, Germany}
s
\author{Morteza Kayyalha}

\affiliation{Department of Electrical Engineering, The Pennsylvania
State University, University Park, Pennsylvania 16802, USA}

\begin{abstract}
Motivated by recent experiments on multiterminal Josephson junctions
(MJJs) that probe different ranges of the size and bias voltage
{parameters, we explore} the regime of increasing bias
voltage in large-scale {devices, where} the
electrochemical potential becomes comparable to the 1D energy level
{spacing. We find} that the relative number of
quantum-correlated pairs formed by colliding Floquet--Kulik
{quartet} levels is equal to the inverse of the number
of {channels. This} observation motivates a model for
the intermediate regime in which the ballistic central two-dimensional
normal metal is treated as a continuum under the adiabatic
approximation, while Andreev modes propagate in a background of
voltage- and flux-tunable nonequilibrium electronic populations. The
model predicts characteristic voltage scales that govern the
mesoscopic oscillations of the critical current, and these scales are
at the crossroads of interpreting experiments in all sectors of the
MJJs: quartets, topology, and Floquet theory. {Our
  model is specifically inspired by the recent Harvard and Penn State
  group experiments.}
\end{abstract}

\maketitle

\section{Introduction}
Multiterminal Josephson junctions (MJJs) have attracted considerable
interest in recent years, both theoretical
\cite{Omelyanchouk0,Omelyanchouk1,Omelyanchouk2,Omelyanchouk3,Omelyanchouk4,Freyn2011,Cuevas-Pothier,Melin2016,Melin-Sotto,Melin2017,topo1,topo2,topo3,topo4,Melin2019,molecule1,molecule3,Melin2020,Melin2020a,Doucot2020,molecule2,Melin2021,Melin2022,Akh2,Melin2023,Melin2023a,Keliri2023a,Keliri2023b,molecule4,LY,Melin2024,Melin2024a,Heck2014,Jonckheere,Rech,Padurariu}
and experimental
\cite{Pfeffer2014,Strambini2016,Cohen2018,Draelos2019,Pankratova2020,Graziano2020,Arnault2021,Huang2022,Graziano2022,Arnault2022,Matsuo2022,Zhang2023,Gupta2023,Matsuo2023a,Matsuo2023b,Coraiola2023,Coraiola2024a,Coraiola2024b,Kurtossy2021,Park2022,Gupta2024,Zhang2024,Rashid2025,Arnault2025,Wisne2024,Prosko2024,Behner2024,Jung2025,Girit2024,Nichele2025},
{driven by the goal of engineering devices that
 exploit the wider range of control parameters made possible by
 multiple superconducting leads.} For instance, in three- and
four-terminal Josephson junctions, {experimental
 studies} \cite{Pfeffer2014,Cohen2018,Huang2022,Gupta2024}
{have confirmed the theoretically predicted
 correlations among Cooper pairs} \cite{Freyn2011}--the so-called
quartets, sextets, octets, and beyond. {The quartets
 are the current-carrying intermediate states involving four
 fermions, that exchange partners among the superconducting
 contacts.} {Predictions of topological effects in
 these systems have also inspired both} theoretical
\cite{topo1,topo2,topo3,topo4,Heck2014,Padurariu}
{and} experimental
\cite{Strambini2016,Wisne2024,Jung2025,Nichele2025}
{investigations.} While zero-energy states
\cite{Heck2014,Padurariu} {have been reported
 experimentally}
\cite{Strambini2016,Wisne2024,Jung2025,Girit2024,Nichele2025},
{a complete analogy with the quantized edge states of
 the quantum Hall effect has yet to be established.} Furthermore,
hybridization between Andreev bound states, {forming
 so-called Andreev molecules, has been studied theoretically}
\cite{molecule1,molecule2,molecule3,molecule4} {and
 observed experimentally}
\cite{Kurtossy2021,Matsuo2023a,Matsuo2023b,Coraiola2023}.

{More recently, Floquet theory in MJJs has attracted
 significant attention}
\cite{Melin2017,Melin2019,Melin2020a,Doucot2020,Keliri2023a,Keliri2023b},
{but experimental signatures of Floquet replicas
 remain largely absent, with the notable exception of Floquet-Andreev
 bound states (Floquet-ABS) observed in graphene-based,
 microwave-irradiated JJs} \cite{Park2022}.

Those observations motivate the present discussion of {low-voltage
  scales in MJJs, particularly in light of the growing experimental
  interest in ballistic two-dimensional (2D) metal-based devices.} The
adiabatic limit is usually understood as a slow evolution of the
superconducting phase variables, {implemented by taking the $V = 0^+$
  limit of infinitesimal voltage in transport formulas.}  {Here, we go
  one step further and highlight an intermediate regime in which the
  electronic populations experience a finite bias voltage $V$, while
  the superconducting phase variables evolve adiabatically and are
  evaluated at $V=0^+$.} This approximation is physically motivated by
the observation that the Floquet theory for large-scale MJJs exhibits
{limited interlevel quantum coherence}.  {Accordingly, we implement a
  model that neglects quantum-mechanical Landau--Zener tunneling in
  the phase dynamics, but retains the nonequilibrium electronic
  populations.} The resulting {simple, physically-motivated} model
serves as a guideline for interpreting experiments on 2D-metal-based
MJJs, {such as the recent Harvard group study}~\cite{Huang2022}.

{Regarding voltage biasing, MJJ studies typically fall
 into two categories: equilibrium at vanishingly small bias voltage,
 and nonequilibrium at finite bias voltage.} In equilibrium
four-terminal JJs, the critical current is defined as a contour in the
plane of the two bias currents—the \textit{critical current
 contour}~\cite{Pankratova2020,Melin2023a,Omelyanchouk0}. At finite bias voltage,
{experimental results indicate the presence of
 intrinsic quantum-mechanical effects beyond the resistively shunted
 Josephson junction (RSJ) model}. For example, {the
 classical RSJ model is inconsistent with the quantum noise
 cross-correlation measurements of the Weizmann
 group}~\cite{Cohen2018}, and {the disappearance of
 the quartet line in devices with spatially separated
 JJs}~\cite{Cohen2018,Pfeffer2014} {cannot be
 explained by quantum synchronization from the electromagnetic
 environment}.

Increasing the bias voltage {can produce a wide range
 of phenomena coupling the DC
 current}~\cite{Freyn2011,Pfeffer2014,Cohen2018,Huang2022,Gupta2024}
{or quantum noise}~\cite{Melin-Sotto,Cohen2018}
{to the spectrum and to the electronic populations}.
Theoretical possibilities include: (i) {emergence of
 an energy-periodic spectrum, as in a Fabry--P\'erot interferometer};
(ii) coupling of signals to nonequilibrium populations induced by
voltage-biased leads; (iii) formation of static quartet--Andreev bound
states (ABS); (iv) their Floquet replicas; (v) nonadiabatic
Landau--Zener transitions; (vi) multiple Andreev reflections (MARs)
connecting low-energy states to quasiparticles above the gaps.
{Solving the full theory for items (i)--(vi) across
 arbitrary device dimensions $L$, separations $R$, transparencies,
 and voltages $V$ is highly challenging.} {Here, we
 focus on identifying experimentally relevant regimes and formulating
 the simplest models.}

In the following, {we argue that} quantum correlations in Floquet
spectra superimposed on a continuum are meaningful in one dimension
but not in two dimensions, where the fraction of correlated state
pairs is inversely proportional to the number of transverse
channels. Assuming a continuous spectrum, we {introduce a model
  describing the interaction between quartet modes and the background
  nonequilibrium electronic populations sustained by the ballistic
  2D-metal density of states.} The populations remain at
nonequilibrium under a small but finite bias, while {the
  superconducting phase variables evolve adiabatically, eliminating
  quantum-mechanical Landau--Zener tunneling and remaining consistent
  with Floquet--Kulik observations}.

{Previous works focusing on the strict adiabatic limit
  for large-scale 2D metal-based MJJs \cite{Melin2020} successfully
  explained the experimental \cite{Huang2022} sensitivity of the
  quartet anomaly as a function of the magnetic field. Conversely, the
  dependence of the quartet anomaly on the bias voltage was addressed
  within the Floquet theory for a zero-dimensional quantum dot
  consisting of a single tight-binding site connected to four
  superconductors \cite{Melin2020a}. The goal of the present work is
  to provide a single theoretical framework to account for both the
  magnetic field and the voltage-dependence of the quartet signal in
  large-scale 2D metal-based MJJs. We specifically enhance the
  adiabatic limit with the addition of a coupling to the
  nonequilibrium electronic populations, via the magnetic flux and
  bias voltage-dependent electrochemical potential of the 2D metal
  nonequilibrium Fermi surface. This approach is further supported by
  a recent experiment \cite{Penn-State-spectro} on the tunneling
  spectroscopy of three-terminal Josephson junctions.}

{The paper is organized as follows. The devices and
  the Hamiltonians are presented in section~\ref{sec:dev+H}. The
  enhanced adiabatic approximation is further justified in
  section~\ref{sec:FK} on the basis of an evaluation of the
  voltage-dependence of the semiclassical spectra of the quartets.
  The results are presented in section~\ref{sec:noneq}. Final remarks
  are provided in section~\ref{sec:conclusions}.}

\begin{figure}[htb]
  \includegraphics[width=.7\columnwidth]{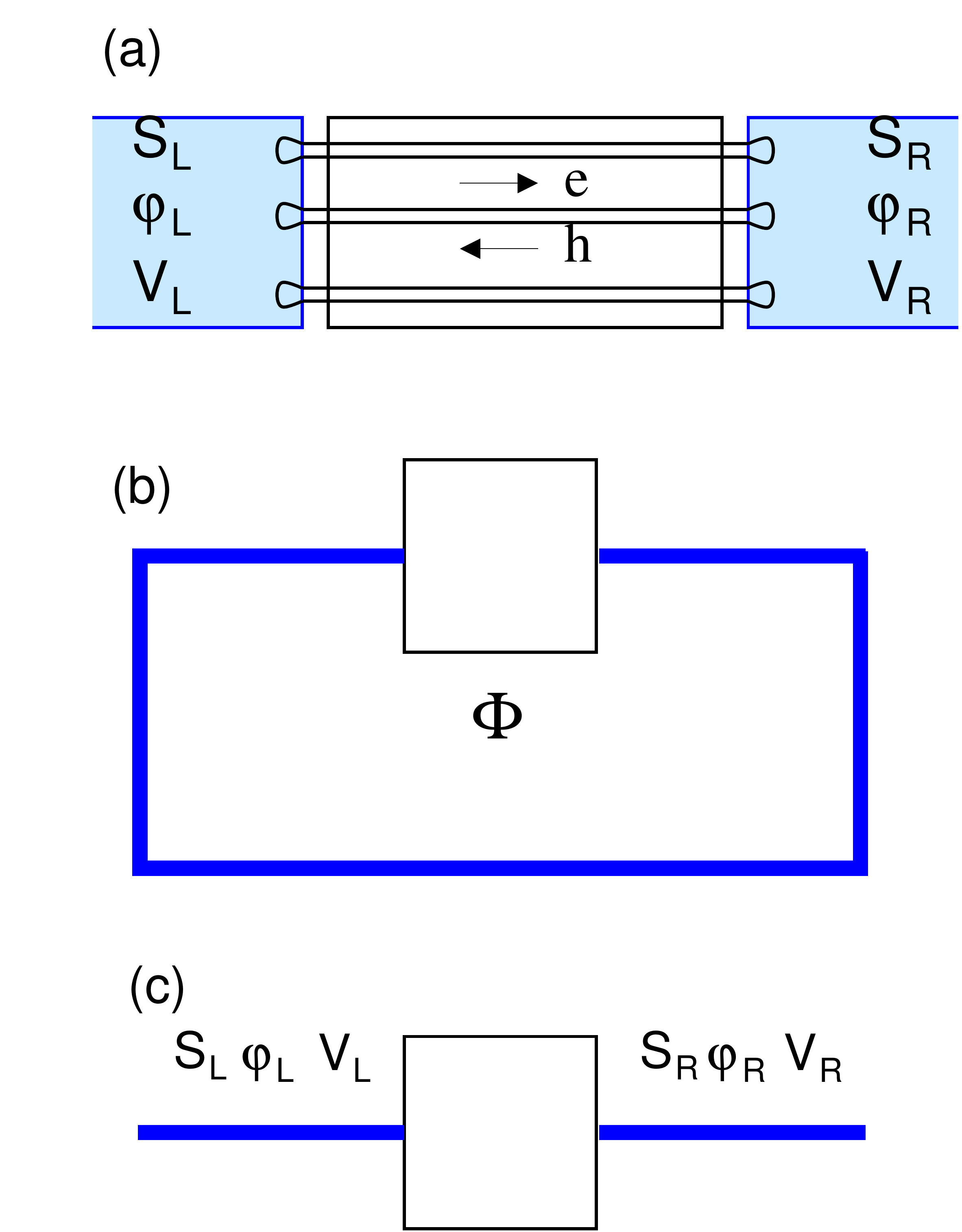}
  \caption{The two-terminal JJ: The horizontal Andreev tubes that
   carry the pairs between the left and right superconducting leads
   (a); the 2D JJ embedded in a superconducting loop (b); and the
   voltage-biased two-terminal JJ (c).
 \label{fig:device-2T}
}
\end{figure}

\begin{figure}[htb]
  \includegraphics[width=.7\columnwidth]{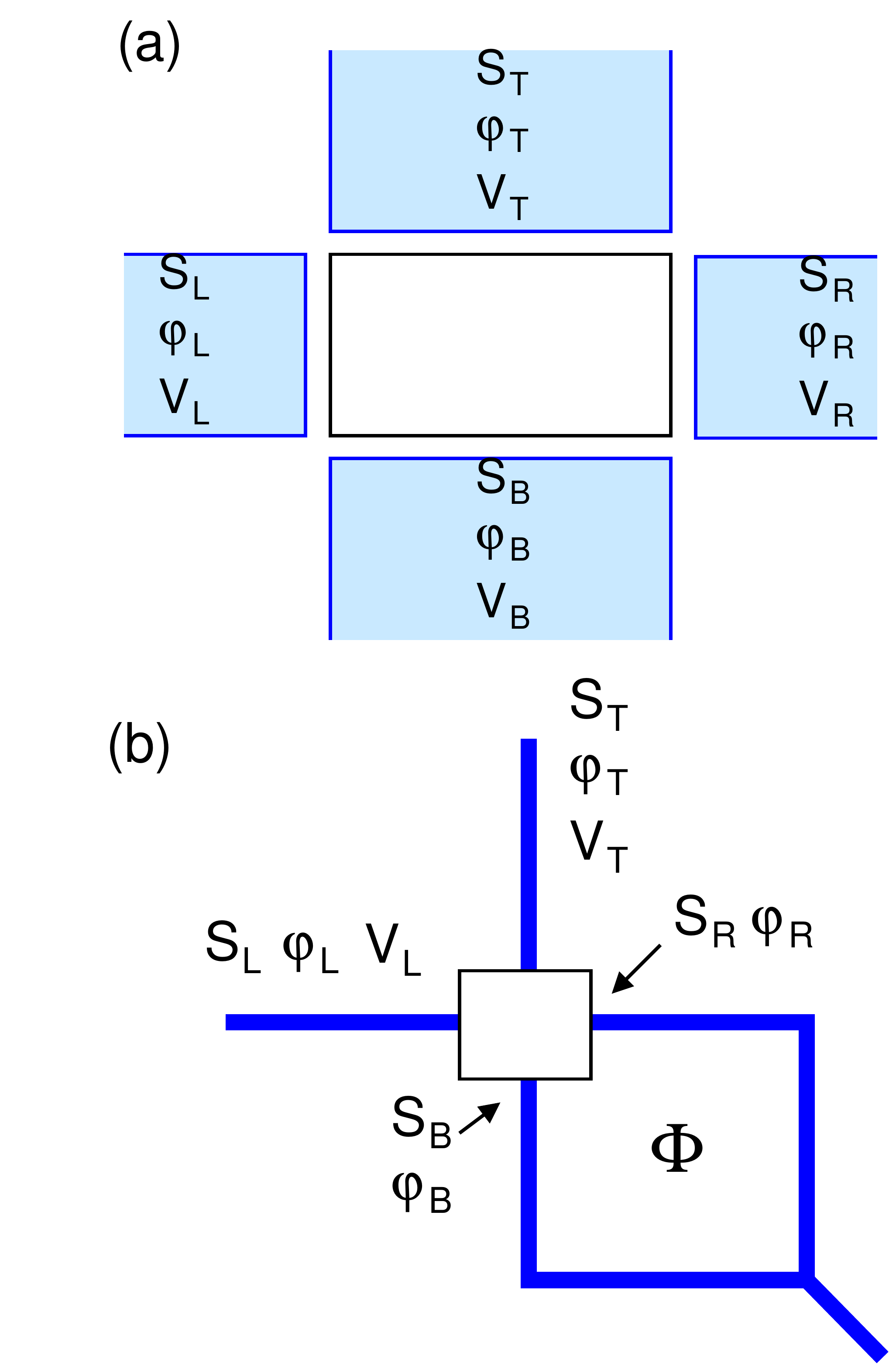}
  \caption{The four-terminal JJ: The device (a) and the circuit containing
   a loop (b).
 \label{fig:device-4T}
}
\end{figure}

\begin{figure}[htb]
  \includegraphics[width=.7\columnwidth]{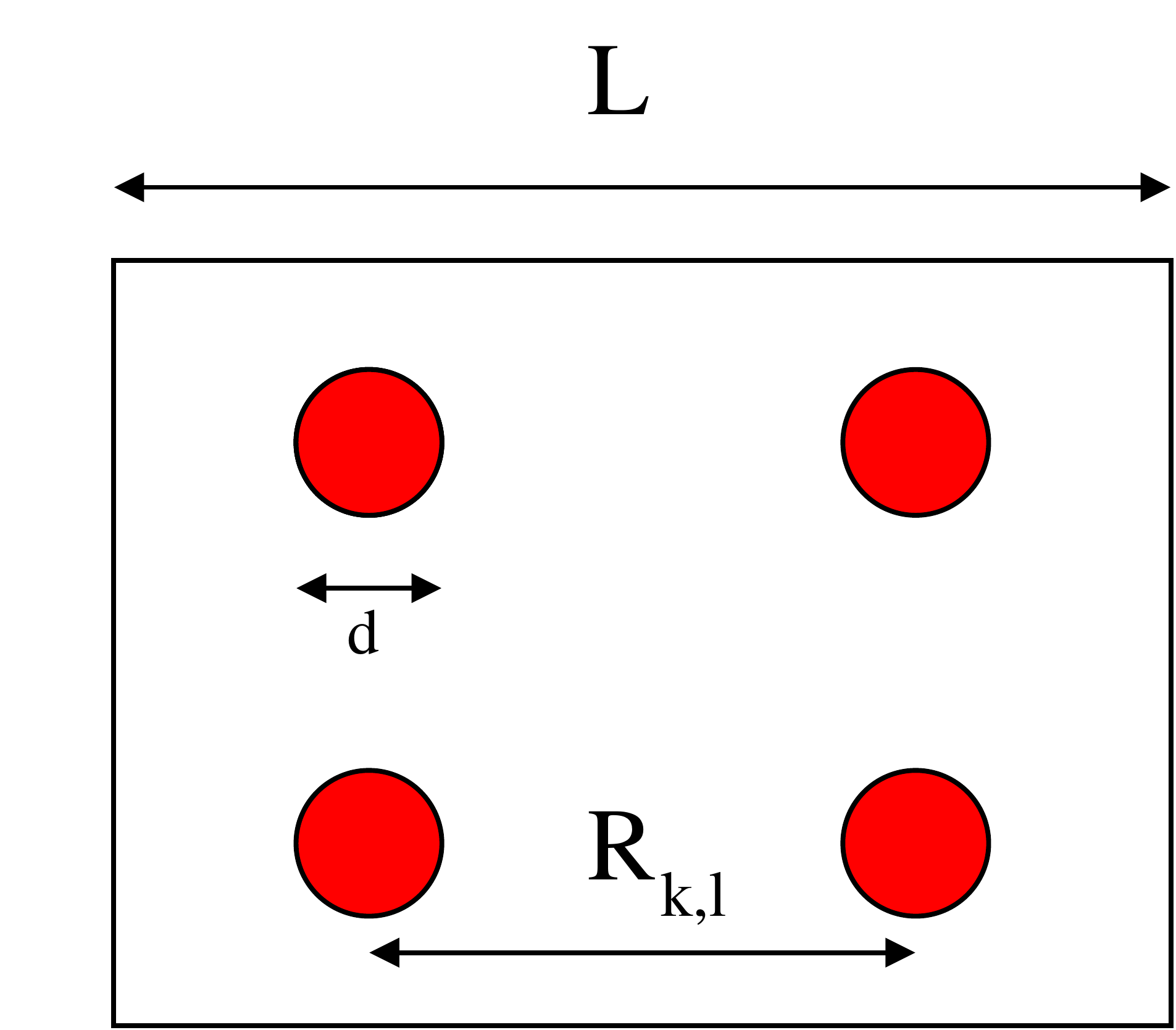}
  \caption{The four-terminal JJ evaporated on top of a ballistic
   2D-metal of dimension $L\times L$. The distance between the
   superconducting contacts $S_k$ and $S_l$ is denoted by
   $R_{k,l}$. The diameter of each contact is denoted by $d$. The
   spectrum is continuous if $L$ is sent to infinity.
 \label{fig:device-4T-RL}
}
\end{figure}
 
\section{Devices and Hamiltonians}
\label{sec:dev+H}

In this introductory section, we present the devices in
subsection~\ref{sec:devices} and provide the Hamiltonians in
subsection~\ref{sec:Hamiltonians}.

\subsection{Devices}

\label{sec:devices}

In this subsection, we introduce the devices that are considered in
the remaining of the paper. First, {we discuss in the
  forthcoming section~\ref{sec:FK} a simple} two-terminal
superconductor-normal metal-superconductor $S_L$-$N$-$S_R$ device
biased at $(V_L,V_R)$, see Fig.~\ref{fig:device-2T}a. The time-$t$
dependence of the corresponding phase variables is the following:
\begin{eqnarray}
 \label{eq:phase-dyn-2T-1}
\varphi_L(t)&=&\varphi_L+{2eV_Lt}/{\hbar}\\
\varphi_R(t)&=&\varphi_R+{2eV_Rt}/{\hbar}
\label{eq:phase-dyn-2T-2}
.
\end{eqnarray}

{We will also consider that the three superconducting terminals $S_L$,
  $S_R$ and $S_B$ are connected to a rectangular 2D-metal, and biased
  at the voltages $(V_L,V_R,0)$. The corresponding superconducting
  phase variables are given by
  Eqs.~(\ref{eq:phase-dyn-2T-1})-(\ref{eq:phase-dyn-2T-2}) and
  $\varphi_B(t)=\varphi_B$.}

We {will} next calculate in the forthcoming
section~\ref{sec:noneq} the phase-sensitive current in the presence of
a continuum, for a four-terminal device. We consider a device where
the four left, right, top and bottom superconducting leads $S_L$,
$S_R$, $S_T$ and $S_B$ are connected to the edges of a rectangular
ballistic 2D-metal, see Fig.~\ref{fig:device-4T}a.

In some of the model calculations, we assume that the 2D-metal has
dimension $L\times L$, see Fig.~\ref{fig:device-4T-RL}. It is assumed
that four superconducting leads $S_k$ are evaporated on the ballistic
2D-metal, and the distance between $S_k$ and $S_l$ is denoted by
$R_{k,l}$. We assume that the contact area scales like $\approx \pi
d^2/4$, where the contact diameter $d$ is such that $d\gg\lambda_F$,
with $\lambda_F$ the ballistic 2D-metal Fermi wave-length. Then,
sending $L$ to infinity according to $L\rightarrow \infty$ produces a
continuous spectrum, keeping constant the separations $R_{k,l}$
between all the pairs of contacts $(S_k,S_l)$.

The superconducting leads $S_L$ and $S_T$ are assumed to be biased on
the quartet line at the opposite voltages $V_{L,T}=\pm V$. The leads
$S_R$ and $S_B$ terminate the large grounded superconducting loop
pierced by the magnetic flux~$\Phi$, see
Fig.~\ref{fig:device-4T}b. The superconducting phase variables of the
corresponding four-terminal device have the following time-$t$
dependence:
\begin{eqnarray}
 \label{eq:phases-1}
 \varphi_L(t)&=&\varphi_L+{2eVt}/{\hbar}\\
 \varphi_R(t)&=&\varphi_R\\
 \varphi_T(t)&=&\varphi_T-{2eVt}/{\hbar}\\
 \varphi_B(t)&=&\varphi_B.
 \label{eq:phases-2}
\end{eqnarray}


\subsection{Hamiltonians}

\label{sec:Hamiltonians}

In this subsection, we present the Hamiltonians that are used
throughout the paper. The Hamiltonian ${\cal H}_{2D,N}$ of the
central-$N$ is given by the following tight-binding model with hopping
amplitude $W$ on a 2D-lattice:
\begin{equation}
 \label{eq:H-tb-2D}
 {\cal H}_{2D,N}=-W \sum_{\langle n,m \rangle} \sum_\sigma
 \left(c_{n,\sigma}^+ c_{m,\sigma} + c_{m,\sigma}^+
 c_{n,\sigma}\right) ,
\end{equation}
where $\langle n,m \rangle$ denotes the pairs of the neighboring
tight-binding sites $n$ and $m$, and $\sigma$ is the projection of the
spin on the quantization axis.

We additionally assume that a well-defined Fermi surface is produced
by gating the ballistic 2D-metal away from singularities such as the
Dirac points or the van Hove singularities. If the bias voltages and
the superconducting gaps are much smaller than the band-width, then
the ballistic 2D-metal dispersion relation associated to
Eq.~(\ref{eq:H-tb-2D}) can be linearized, and it can be parameterized
by the two numbers of the Fermi energy~$\epsilon_F$ and the Fermi
velocity~$v_F$. The Hamiltonian ${\cal H}_g$ associated to the gate
voltage $V_g$ is defined as
\begin{equation}
 {\cal H}_g=-W_g \sum_n \sum_\sigma
 c_{n,\sigma}^+ c_{n,\sigma}
 ,
\end{equation}
where the energy $W_g$ is proportional to the voltage $V_g$.

Each superconducting lead is described by the standard BCS Hamiltonian
${\cal H}_{BCS}$ with the order parameter $\Delta_n \exp(i\varphi_n)$
at the tight-binding site labeled by $n$, where $\Delta_n$ and
$\varphi_n$ are the superconducting order parameter amplitude and the
superconducting phase, respectively:
\begin{eqnarray}
 \label{eq:H-BCS}
&& {\cal H}_{BCS} = - W \sum_{\langle n,m\rangle} \sum_\sigma
 \left( c_{n,\sigma}^+ c_{m,\sigma} + c_{m,\sigma}^+ c_{n,\sigma}\right)\\
 &-&\sum_n \left( \Delta_n
 \exp(i\varphi_n) c_{n,\uparrow}^+ c_{n,\downarrow}^+ +
 \Delta_n \exp(-i\varphi_n) c_{n,\downarrow} c_{n,\uparrow}
 \right)
 ,
 \nonumber
\end{eqnarray}
where we assume in Eq.~(\ref{eq:H-BCS}) the same hopping amplitude as
in Eq.~(\ref{eq:H-tb-2D}) for the ballistic 2D-metal.

The coupling between the interfaces is via the hopping Hamiltonian
${\cal H}_{tun}$ with the hopping amplitude $\Sigma_0$:
\begin{equation}
 {\cal H}_{tun}= - \Sigma_0 \sum_{n_1,n_2} \sum_\sigma \left(
 c_{n_1,\sigma}^+ c_{n_2,\sigma} + c_{n_2,\sigma}^+
 c_{n_1,\sigma}\right) ,
\end{equation}
where the integers $n_1$ and $n_2$ label the corresponding
tight-binding sites on both sides of the interface. The tunnel
contacts correspond to $\Sigma_0\ll W$ and the highly transparent
interfaces are such that $\Sigma_0=W$.

The coupling to the magnetic flux $\Phi$ piercing through the loop in the
four-terminal device of Fig.~\ref{fig:device-4T}b is via the
following transformation for the choice of the gauge:
\begin{eqnarray}
 \label{eq:gauge1}
 \varphi_L&\rightarrow&\varphi_L\\ \varphi_R&\rightarrow&
 \varphi_R+\frac{2\pi\Phi}{\Phi_0}\\ \varphi_T&\rightarrow&
 \varphi_T\\ \varphi_B &\rightarrow& \varphi_B ,
 \label{eq:gauge4}
\end{eqnarray}
The flux $\Phi$ is calculated from the line integral of the vector
potential along a closed contour encircling the superconducting loop
and continuing through the ballistic 2D-metal. The contribution of the
ballistic 2D-metal to the line integral of the vector potential is
neglected with respect to the contribution of the loop that is assumed
to have much larger dimension.

{\section{Collisions between the Floquet--Kulik quartet superlevels}\label{sec:FK}}

In this section, we provide a simple argument resolving the collision
between the Floquet--Kulik superlevels {of the
  quartets}, if the spectrum is plotted as a function of the bias
voltage in a long JJ connected by highly transparent interfaces
\cite{Kulik,Ishii,Bagwell}.

The spectrum of long, highly transparent and ballistic $S$-$N$-$S$
junctions was calculated in a wave-guide geometry by Kulik
\cite{Kulik}, and further discussed by Ishii \cite{Ishii} and Bagwell
\cite{Bagwell}. In the ideal wave-guide geometry of a rectangle of
dimension $N a_0 \times M a_0$, the Kulik spectrum consists of $M$
degenerate ladders, where $M$ is the number of transverse channels
{and $a_0$ is the lattice spacing.} Physically, the
supercurrent is carried by independent Andreev tubes of transverse
dimension set by $\lambda_F$, and traversing the central-$N$ between
the left and the right sides, parallel to the $x$-axis, see
Fig.~\ref{fig:device-2T}a.

A discrete spectrum of the Floquet states is produced by voltage
biasing MJJs, due to time-periodic Hamiltonian. The Floquet spectrum
of zero-dimensional quantum dots connected to superconducting leads
was calculated in Refs.~\onlinecite{Melin2017,Melin2019}.

In the presence of voltage-biasing, the Floquet Hamiltonian is
block-diagonal on the basis of the Andreev tubes in real space.  We
first argue that a long $S_L$-$N$-$S_R$ Josephson junction subject to
the voltage difference $V_R-V_L$, and with the phase dynamics of
{Eqs.~(\ref{eq:phase-dyn-2T-1})-(\ref{eq:phase-dyn-2T-2}),
  does not sustain a Floquet spectrum. Namely, the AC-Josephson effect
  is represented by a diagram that is open in the energy
  representation, which produces absence of closed semiclassical
  trajectories. Instead, the diagrams of multiple Andreev reflections
  (MARs) form closed loops but inspecting those diagrams yields the
  absence of energy-dependence in the limit of an infinite gap and the
  absence of semiclassical energy quantization in the limit of a long
  junction.}

{Concerning the quartets, the butterfly diagram \cite{Freyn2011} can
  be used to sustain the closed trajectories that are used to
  implement the Bohr-Sommerfeld semiclassical quantization. We deduce
  that a Kulik spectrum is there for the quartets in the
  voltage-biased three- and four-terminal Josephson junctions.}
{We then obtain collisions between the $M$-fold
  degenerate Floquet--Kulik superlevels. Generalizing the calculations
  of the recent Ref.~\onlinecite{Penn-State-spectro} to finite bias
  voltage $V$ on the quartet line, we find the following semiclassical
  energy levels in three-terminal Josephson junctions:
  \begin{equation}
    \label{eq:Omega-pm}
    \Omega^{(\pm)}_{n_{(\pm)}}(eV,\chi_q)
    = \pm eV +
    \frac{\hbar v_F}{2(R_{\alpha,\beta}+R_{\beta,\gamma})}
    \left(\pm \chi_q + 2 \pi n_{(\pm)}\right)
    ,
  \end{equation}
  where $v_F$ denotes the normal metal Fermi velocity, $\chi_q$ is the
  quartet phase $\chi_q=\varphi_L+\varphi_R-2\varphi_B$, where
  $\varphi_L$, $\varphi_R$ and $\varphi_B$ are the superconducting
  phase variables of the left, right and bottom superconducting leads
  $S_L$, $S_R$ and $S_B$. The contacts between the superconductors
  $S_L$, $S_R$, $S_B$ biased at $V_L=V$, $V_R=-V$ and $V_B=0$, and the
  central 2D metal $N$ are denoted by $(a,\,\alpha)$, $(b,\,\beta)$
  and $(c,\,\gamma)$, where $a$, $b$ and $c$ are in $S_L$, $S_R$ and
  $S_B$ and $\alpha$, $\beta$ and $\gamma$ are their counterparts in
  $N$. The notations $R_{\alpha,\beta}$ and $R_{\beta,\gamma}$ stand
  for the separations between the $\alpha$-$\beta$ and
  $\beta$-$\gamma$ pairs of tight-binding sites in $N$. The two
  variables $n_{(\pm)}$ are integers that label the Kulik states of
  the electron-like and hole-like types. Solving for
  $\Omega^{(+)}_{n_{(+)}}(eV^*,\chi_q)=\Omega^{(-)}_{n_{(-)}}(eV^*,\chi_q)$
  leads to the voltage value $V^*(n_{(+)},n_{(-)}$ of the collision
  between the two families of the Floquet-Kulik states, which move
  upwards or downwards as a function of the voltage $V$.}

Considering a collision between two of those $M$-fold degenerate
Floquet--Kulik quartet superlevels, we note that only $M$ pairs of
Andreev tubes are quantum mechanically coupled, due to a kind of
fractionalization of the Hilbert space resulting from the 1D geometry
of the Andreev tubes. We conclude that the conservation laws of those
intra or intertube level crossings are such that, in 2D, $M$ pairs of
the constitutive Floquet-ABS are quantum mechanically correlated and
$M^2-M$ pairs are quantum mechanically uncorrelated. This argument
indicates that the Floquet--Kulik superlevel collisions have the mixed
features of the coexistence between the nonavoided and avoided
crossings. The relative amount of the quantum tunneling in the
Floquet--Kulik spectra {of the quartets} thus scales
like $1/M$ in the limit of a large number $M$ of the transverse
channels. We conclude to a kind of {\it dilution} of the quantum
mechanical correlations of Landau-Zener tunneling in the large-$M$
limit of the 2D-devices, i.e. the relative weight of the Landau-Zener
quantum tunneling goes to zero in the large-$M$ limit.

\section{Forming an interference in the phase-sensitive nonequilibrium current}
\label{sec:noneq}

{We discussed} in the above section~\ref{sec:FK} that,
in 2D, the collisions between the Floquet--Kulik
{quartet} superlevels produce $M$ and $M^2-M$ pairs of
quantum correlated and quantum uncorrelated Floquet--Kulik energy
levels, respectively{, where $M$ is the number of the transverse
  channels.} A fraction $x=1/M$ of the pairs of the colliding
Floquet-Kulik {quartet} levels is thus sensitive to
the energy level repulsion of the quantum mechanical Landau-Zener
tunneling. Those quantum mechanical correlations have thus probability
that scales like $x=1/M$. Given that $x\rightarrow 0$ as $M$ is
increased to reach a large-scale 2D-device, we now implement a
classical approximation for the phase dynamics in large-scale devices,
which results in the complete absence of the Landau-Zener
tunneling. Namely, the time-periodic dynamics is simply treated as
being adiabatic (i.e. the $V=0^+$ limit is implemented). We
additionally assume the electron-phonon relaxation towards a
nonequilibrium Fermi surface with the electrochemical potential
$\mu_N$. The shift in the electrochemical potential
$\delta\mu_N\left(eV, \frac{\Phi}{\Phi_0}\right)$ (with respect to its
equilibrium value) is assumed to be sensitive to both the bias voltage
$V$ and to the magnetic flux $\Phi$, see
Refs.~\onlinecite{Stoof-Nazarov1,Stoof-Nazarov2}.

{We detail in Appendix~\ref{sec:known}} the
microscopic ingredients that sustain the calculation of the
current-phase relations in the limiting case of small-transparency
interfaces. {In the main text, we discuss} the
phenomenology of the interference between those different types of
quartet current-phase relations. {The even-in-phase
  current-phase relations are presented in
  subsection~\ref{sec:even-in-phase} and the numerical results are
  presented in subsection~\ref{sec:num-inter}.}

\subsection{Even-in-phase current-phase relation}
\label{sec:even-in-phase}

In this subsection, we calculate the quartet current-phase
relations in the presence of a coupling to the background
nonequilibrium electronic populations. The latter are produced in the
central-$N$ simply by voltage-biasing in the presence of
nonsymmetrical contact transparencies. We implement the simplifying
assumption of strong electron-phonon coupling, which produces
step-function energy-variations in the distribution functions, having
the single-step of weight $Z=1$ at the electrochemical potential
$\delta\mu_N\left(e(V,\frac{\Phi}{\Phi_0}\right)$. It is demonstrated
in Appendix~\ref{sec:sq-long} that those low-lying quartet ABS yield
the current-phase relation $I_q= I_{q,3T,c}^{(eff)}[\delta \mu_N]
\cos\varphi_q$, where the critical current $I_{q,3T,c}^{(eff)}$
oscillates as a function of the electrochemical potential:
\begin{eqnarray}
 \label{eq:Iqceff}
 && I_{q,3T,c}^{(eff)}[\delta \mu_N]=\\ \nonumber
 &&I_{q,3T,c}^{(even)}\times \delta\mu_N \times
 \cos\left(\frac{2\delta \mu_N R_{\alpha,\gamma}}{\hbar v_F}\right)
 \cos\left(\frac{2\delta \mu_N R_{\beta,\gamma'}}{\hbar v_F}\right)
\end{eqnarray}
scales like $I_{q,3T,c}^{(eff)}[\delta \mu_N] \simeq
I_{q,3T,c}^{(even)} \times \delta \mu_N$ at small ${2\delta \mu_N
 R_{\alpha,\gamma}}/{\hbar v_F}\alt 1$ and ${2\delta \mu_N
 R_{\beta,\gamma'}}/{\hbar v_F} \alt 1$, and oscillates as $\delta
\mu_N$ is increased. In those expressions, we assumed that $\delta
\mu_N$ is small compared to the superconducting gap $\Delta$,
i.e. $\delta\mu_N\ll\Delta$. The growth of $I_{q,3T,c}^{(even)}$ with
$\delta\mu_N$, see $I_{q,3T,c}^{(even)}\propto \delta\mu_N$ in
Eq.~(\ref{eq:Iqceff}), is expected to be limited as $\delta \mu_N$
reaches a significant fraction of the superconducting gap $\Delta$.

\begin{figure*}[htb]
 \begin{minipage}{.6\textwidth}
  \begin{minipage}{.49\textwidth}
   \includegraphics[width=1.1\textwidth]{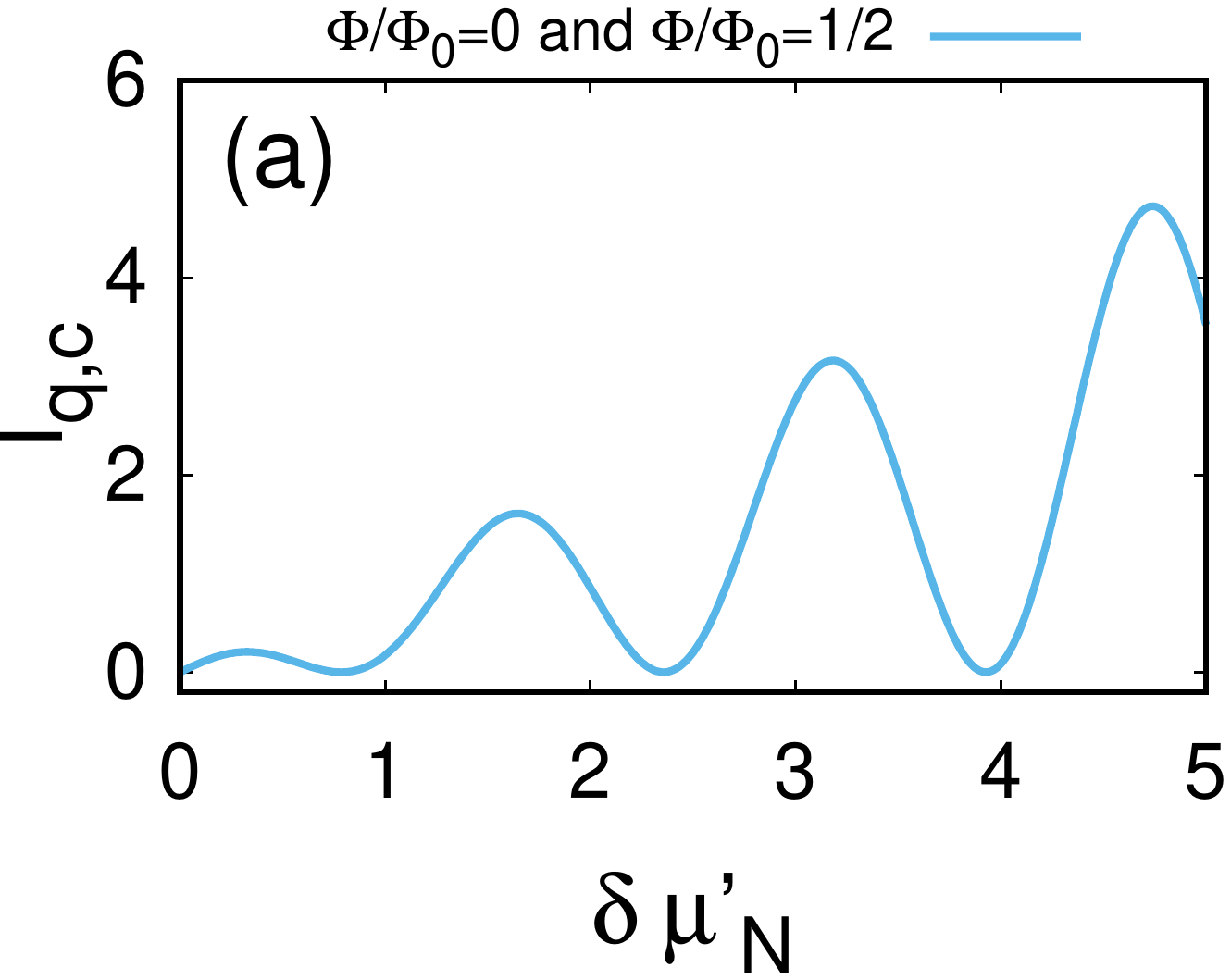}
  \end{minipage}\begin{minipage}{.49\textwidth}
   \includegraphics[width=1.1\textwidth]{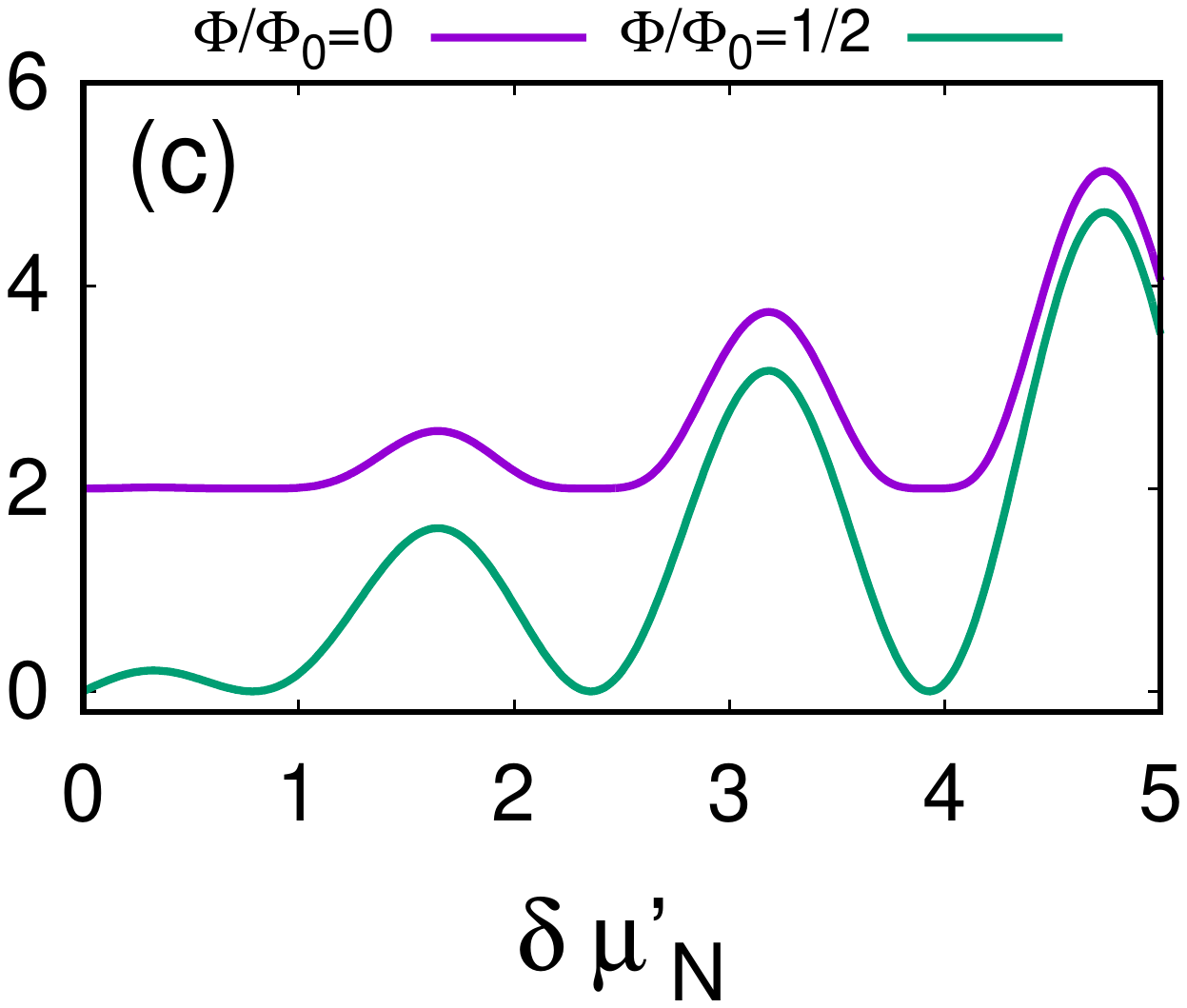}
  \end{minipage}
  \begin{minipage}{.49\textwidth}
   \includegraphics[width=1.1\textwidth]{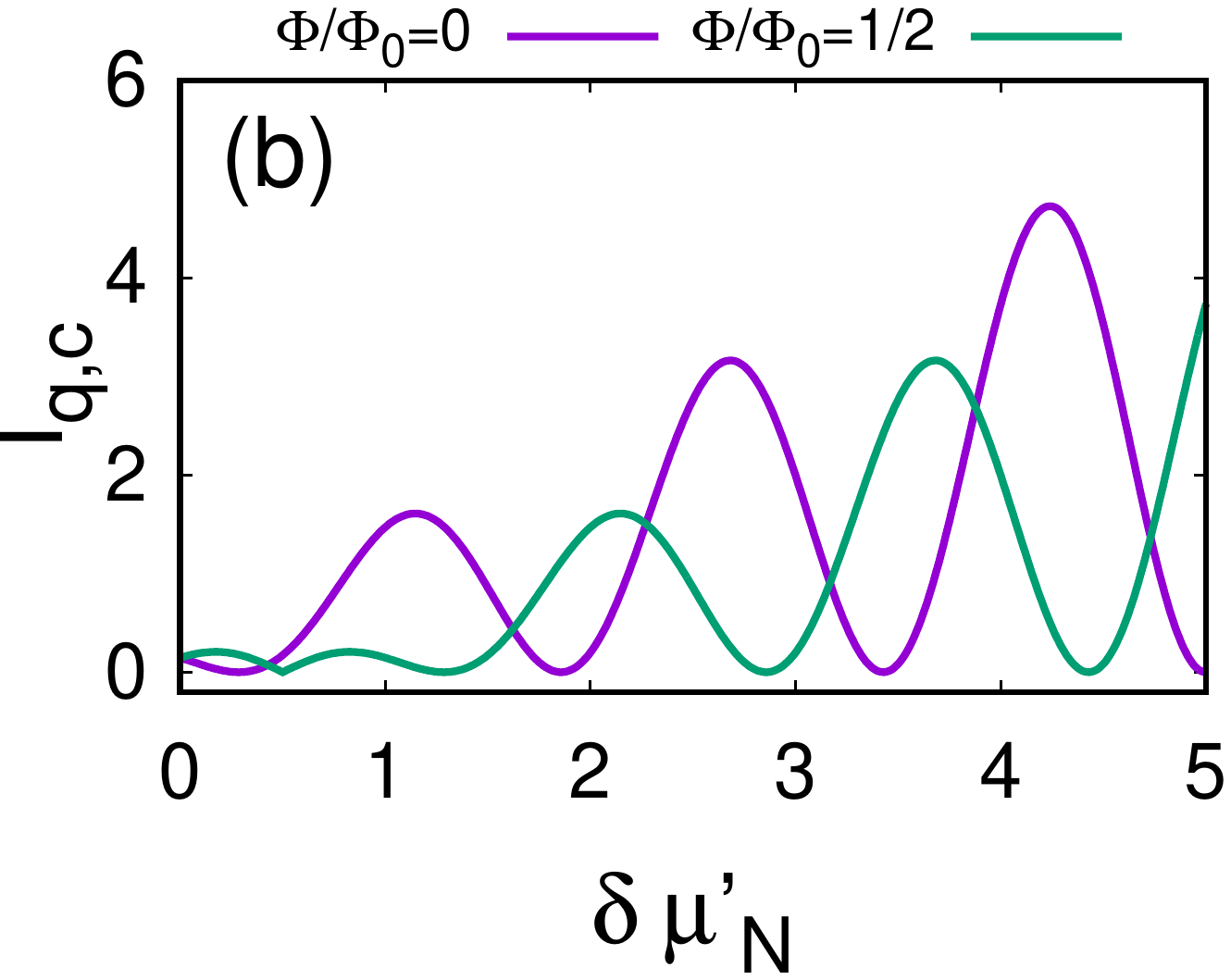}
  \end{minipage}\begin{minipage}{.49\textwidth}
   \includegraphics[width=1.1\textwidth]{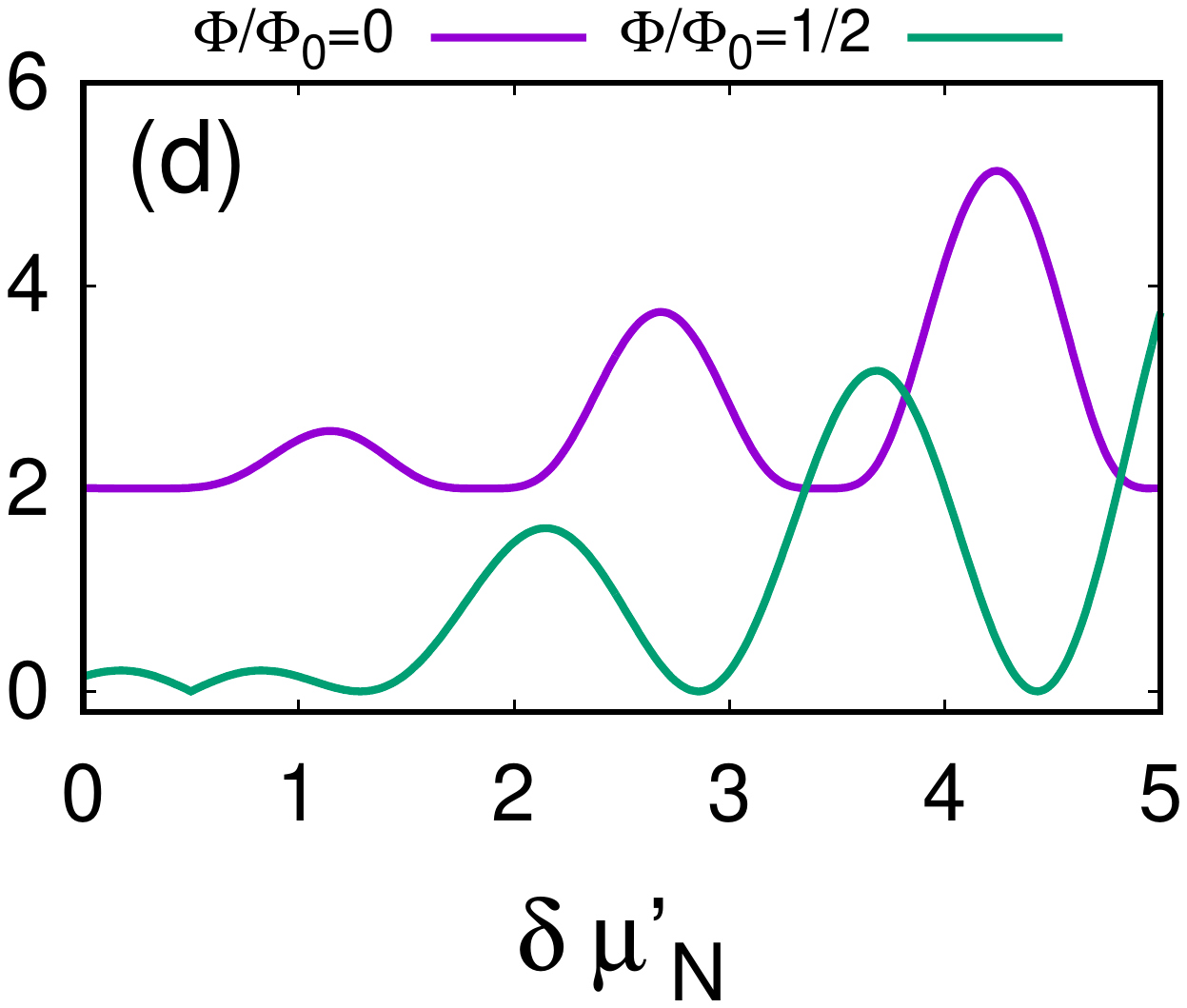}
  \end{minipage} 
 \end{minipage}\begin{minipage}{.39\textwidth}
\caption{The interference $I_{q,c}\left(\Phi,\delta\mu_N\right)$ in
 the critical current between the quartets emitted by the contacts
 $S_R$ and $S_B$, in the presence of the nonequilibrium quasiparticle
 populations in the central-$N$. The magnetic fluxes are
 $\frac{\Phi}{\Phi_0}=0$ (magenta) and
 $\frac{\Phi}{\Phi_0}=\frac{1}{2}$ (green). Panel a shows the same
 data as Fig.~\ref{fig:interference3}, where the critical currents
 are identical for $\frac{\Phi}{\Phi_0}=0$ and for
 $\frac{\Phi}{\Phi_0}=\frac{1}{2}$. Panels c and d show the
 interference $I'_{q,c}\left(\Phi,\delta\mu_N\right)$ between all
 types of quartets emitted by the contacts $S_R$ and $S_B$, see
 Eq.~(\ref{eq:Iqc-4}). Panels b and d include the electroflux effect
 with the transformation $\delta \mu_N\rightarrow \delta \mu_N'\pm
 s_0$, see Eqs.~(\ref{eq:s0-1})-(\ref{eq:s0-2}). Going from panels a
 and b to panels c and d, the electroflux effect qualitatively shifts
 in opposite directions the $\frac{\Phi}{\Phi_0}=0$ and the
 $\frac{\Phi}{\Phi_0}=\frac{1}{2}$ data, which produces the
 bias-voltage sensitive noninversion-inversion cross-overs.
\label{fig:figure_final}
}\end{minipage}
\end{figure*}

\subsection{Numerical results}
\label{sec:num-inter}

In this subsection, we present numerical calculations that
illustrate the above theory. We first assume that the critical current
results from the following interference between both types of the
even-in-phase quartets emitted by the contacts $S_R$ or by $S_B$ at
the extremities of the grounded superconducting loop:
\begin{eqnarray}
 \label{eq:Iqc-3}
 && I_{q,c}\left(\Phi,\delta \mu_N\right)= \mbox{Max}_{\varphi_q}
 \left\{\left[
  i_{R}+i_B\right]\left(\varphi_q,\Phi,\delta\mu_N\right)\right\} ,
\end{eqnarray}
where the components $i_{R,B}\left(\varphi_q, \Phi,
\delta\mu_N\right)$ are deduced from Eq.~(\ref{eq:Iqceff}) encoding
the even-in-phase quartets. Eq.~(\ref{eq:Iqc-3}) produces a
checkerboard pattern of the coherent oscillations in the critical
current, see Fig.~\ref{fig:interference3}. In addition,
Fig.~\ref{fig:interference3} features the identical quartet critical
currents at $\frac{\Phi}{\Phi_0}=0$ and
$\frac{\Phi}{\Phi_0}=\frac{1}{2}$, which is in agreement with the
previously considered interference between the odd-in-phase quartets,
see section~\ref{sec:3Tq}.

We now argue that the experiments such as Ref.~\onlinecite{Huang2022}
are in the intermediate regime of the cross-overs between the discrete
and continuous spectra, and between the short and the long-junction
limits. Therefore, all of the corresponding ABS phenomenologically
contribute to the interference in the supercurrent:
\begin{eqnarray}
 \label{eq:Iqc-4}
&& I'_{q,c}\left(\Phi,\delta\mu_N\right)=
 \mbox{Max}_{\varphi_q}\left\{I_{q,3T,c,R}^{(odd)}
 \sin\left(\varphi_q-\frac{4\pi\Phi}{\Phi_0}\right)\right.\\\nonumber&&
 + I_{q,3T,c,B}^{(odd)} \sin\varphi_q +I_{sq,4T,c}^{(odd)}
 \sin\left(\varphi_q-\frac{2\pi\Phi}{\Phi_0}\right)\\\nonumber &&+
 \left. i_{c,R}
  \cos\left(\varphi_q-\frac{4\pi\Phi}{\Phi_0}\right)
  +
  i_{c,L}\cos\varphi_q \right\}
 .\nonumber
\end{eqnarray}
The first three terms in Eq.~(\ref{eq:Iqc-4}) originate from the ABS
that are sustained by the superconducting gaps in the short-junction
limit, see
Eqs.~(\ref{eq:quartets-short-1})-(\ref{eq:quartets-short-2}) for the
quartets, and Eq.~(\ref{eq:sq-short}) for the split quartets. The
last two terms in Eq.~(\ref{eq:Iqc-4}) reflect the coupling between
the ABS and the nonequilibrium electronic populations.

Finally, we also phenomenologically take into account the possibility
that the electrochemical potential $\delta
\mu_N\left(eV,\frac{\Phi}{\Phi_0}\right)$ is different at the magnetic
fluxes $\frac{\Phi}{\Phi_0}=0$ and
$\frac{\Phi}{\Phi_0}=\frac{1}{2}$. {The electroflux
  effect \cite{Stoof-Nazarov1,Stoof-Nazarov2} corresponds to the
  sensitivity of the electrochemical potential on the flux in the
  loop.} We use the notations
\begin{eqnarray}
 \label{eq:s0-1}
 \delta\mu_N\left(eV,\frac{\Phi}{\Phi_0}=0\right)&=&\delta\mu_N'-s_0\\
 \label{eq:s0-2}
\delta\mu_N\left(eV,\frac{\Phi}{\Phi_0}=\frac{1}{2}\right)&=&\delta
\mu_N' + s_0 .\end{eqnarray} The critical currents are then
parameterized by $\delta \mu_N'$, and they also depend on the strength
$s_0$ of the electroflux effect \cite{Stoof-Nazarov1,Stoof-Nazarov2}.
The shift $\delta \mu_N' \propto V$ in the electrochemical potential
is proportional to the bias voltage $V$ in the presence of
nonsymmetrical contacts between the superconducting leads and the
ballistic 2D-metal.

{For the considered four-terminal Josephson junction
  in Fig.~\ref{fig:device-4T}, the electroflux effect microscopically
  originates from the process of phase-Andreev reflection (phase-AR)
  between the two grounded superconducting leads that terminate the
  superconducting loop, see Ref.~\onlinecite{Melin2024}. Assuming
  small electrochemical potential $\mu_N$, the corresponding phase-AR
  current is given by $I_{phase-AR} =- G^{phase-AR} \mu_N \cos\Phi$
  . The ``local'' Andreev reflection currents are given by
  $I^{AR}_{loc,L} = G^{AR}_L(V_L-\mu_N)$,
  $I^{AR}_{loc,T}=G^{AR}_T(V_T-\mu_N)$, $I^{AR}_{loc,B}=-G^{AR}_B
  \mu_N$, and $I^{AR}_{loc,R}=-G^{AR}_R\mu_N$. The absence of net
  current entering the central $N$ is expressed as
  \begin{equation}
    I_{phase-AR}+I^{AR}_{loc,L}+I^{AR}_{loc,T}+I^{AR}_{loc,B}
    +I^{AR}_{loc,R}=0
    .
  \end{equation}
  The resulting value $\mu_N^*$ of the self-consistent $\mu_N$ is the
  following:
  \begin{equation}
    \label{eq:muN*}
    \mu_N^*(\Phi)=\frac{G_L^{AR} V_L+G_T^{AR} V_T}{G_L^{AR} + G_T^{AR}
      + G_B^{AR} + G_R^{AR} + G^{phase-AR} \cos\Phi} ,
  \end{equation}
  which leads to the above Eqs.~(\ref{eq:s0-1})-(\ref{eq:s0-2}) for
  the electroflux effect. The value of $\delta \mu'_N$ in
  Eqs.~(\ref{eq:s0-1})-(\ref{eq:s0-2}) can be monitored by the bias
  voltage $V_t$ on an additional superconducting lead $S_t$ connected
  by a noninvasive tunnel contact. The direct detection of the central
  2D-metal electrochemical potential and populations, see
  Ref.~\onlinecite{Saclay}, can be performed by connecting a second
  superconducting tunneling probe to the 2D metal.}
  
{We now discuss our main result in connection with
  what we call the noninversion or the inversion. By {\it
    noninversion}, we mean that the quartet critical current is larger
  in zero field $\frac{\Phi}{\Phi_0}=0$ than at half flux-quantum
  $\frac{\Phi}{\Phi_0}=\frac{1}{2}$. The noninversion corresponds to
  the expectation of constructive interference in zero field and
  destructive interference at half-flux quantum. By {\it inversion},
  we mean the nonstandard behavior that the quartet critical current
  is larger at half-flux quantum $\frac{\Phi}{\Phi_0}=\frac{1}{2}$
  than in zero field $\frac{\Phi}{\Phi_0}=0$. The Harvard group
  experiment \cite{Huang2022} reports a cross-over between the
  noninversion and the inversion as the bias voltage is increased
  along the quartet line in a four-terminal Josephson junction. This
  experimental observation is explained by the following calculation.}
  
Fig.~\ref{fig:figure_final} summarizes the sensitivity of the quartet
critical current $I_{q,c}(\delta\mu_N',\frac{\Phi}{\Phi_0})$ on the
value $\delta \mu_N'$ of the electrochemical potential, in zero field
$\frac{\Phi}{\Phi_0}=0$ (magenta) and at half-flux quantum
$\frac{\Phi}{\Phi_0}=\frac{1}{2}$ (green). The different panels of
Fig.~\ref{fig:figure_final} correspond to different assumptions about
the parameters in the calculation of the critical current. In
Fig.~\ref{fig:figure_final}a, the quartet critical current
$I_{q,3T,c}=1$ is taken as the reference, in the absence of the
split-quartets $I_{sq,c}=0$ and also in the absence of the electroflux
effect $s_0=0$, see Eq.~(\ref{eq:Iqc-4}) for the notations
$I_{q,3T,c}$ and $I_{sq,c}$ and Eqs.~(\ref{eq:s0-1})-(\ref{eq:s0-2})
for the notation $s_0$. The resulting critical current
$I_{q,c}(\delta\mu_N',\Phi)$ on panel a is identical to
Fig.~\ref{fig:interference3}, see also Eq.~(\ref{eq:Iqc-3}), and it
takes the same value in zero field $\frac{\Phi}{\Phi_0}=0$ and at half
flux-quantum $\frac{\Phi}{\Phi_0}=\frac{1}{2}$. This is why the same
colorcode is used on Fig.~\ref{fig:figure_final}a for
$\frac{\Phi}{\Phi_0}=0$ and for
$\frac{\Phi}{\Phi_0}=\frac{1}{2}$. This panel confirms the
oscillations in the quartet critical current as $\delta \mu_N'$ is
increased, or, equivalently, as the bias voltage $V$ is
increased. Fig.~\ref{fig:figure_final}c now shows the full
Eq.~(\ref{eq:Iqc-4}) with $I_{q,3T,c}=1$ and $i_{c,R,L}=0.5$, still
with $I_{sq,c}=0$. This panel c reveals that
$I_{q,c}(\delta\mu_N',\frac{\Phi}{\Phi_0}=0) \ne
I_{q,c}(\delta\mu_N',\frac{\Phi}{\Phi_0}=\frac{1}{2})$, and that
$I_{q,c}(\delta\mu_N',\frac{\Phi}{\Phi_0}=0)$ is larger than
$I_{q,c}(\delta\mu_N',\frac{\Phi}{\Phi_0}=\frac{1}{2})$ in the entire
range of $\delta\mu_N'$, corresponding to the {\it
 noninversion}. Figs.~\ref{fig:figure_final}b and d are calculated
with the same parameters as Figs.~\ref{fig:figure_final}a and b, but
now with the additional electroflux amplitude $s_0=0.5$, see
Eqs.~(\ref{eq:s0-1})-(\ref{eq:s0-2}) for the notation
$s_0$. Figs.~\ref{fig:figure_final}b and d reveal the $\delta
\mu_N'$-sensitive alternations between the noninversion
$\left[\right.$i.e. $I_{q,c}(\delta\mu_N',\frac{\Phi}{\Phi_0}=0) >
 I_{q,c}(\delta\mu_N',\frac{\Phi}{\Phi_0}=\frac{1}{2})$$\left.\right]$
and the inversion $\left[\right.$i.e.
 $I_{q,c}(\delta\mu_N',\frac{\Phi}{\Phi_0}=0) <
 I_{q,c}(\delta\mu_N',\frac{\Phi}{\Phi_0}=\frac{1}{2})$$\left.\right]$.
Qualitatively, the $\frac{\Phi}{\Phi_0}=0$ and the
$\frac{\Phi}{\Phi_0}= \frac{1}{2}$ data are shifted in opposite
directions by the electroflux effect corresponding to a finite $s_0$
in Eqs.~(\ref{eq:s0-1})-(\ref{eq:s0-2}). Those opposite shifts produce
crossings in the $\delta \mu'_N$ dependence of the critical current
for $\frac{\Phi}{\Phi_0}=0$ and $\frac{\Phi}{\Phi_0}=\frac{1}{2}$,
which results in possible multiple cross-overs between the noninversion
(i.e. the critical current is larger at $\frac{\Phi}{\Phi_0}=0$ that
at $\frac{\Phi}{\Phi_0}=\frac{1}{2}$) and the inversion (i.e. the
critical current is smaller at $\frac{\Phi}{\Phi_0}=0$ than at
$\frac{\Phi}{\Phi_0}=\frac{1}{2}$).

\section{Conclusions}
\label{sec:conclusions}

{In summary, we have presented a phenomenological
 approach to understanding the emergence of characteristic low-bias
 voltages in intermediate-scale MJJs.} The mesoscopic energy
associated with the contact separation {was taken to
 be} comparable to the nonequilibrium electrochemical potential. We
argued that, in 2D devices, {quantum-mechanical
 Landau--Zener tunneling involves a relative number of Floquet--Kulik
 state pairs that scales inversely with the number of transverse
 channels.} {On this basis, we proposed a model
 within the adiabatic approximation, in which Landau--Zener tunneling
 is neglected but nonequilibrium electronic populations in the
 central ballistic 2D metal are retained.} We {used
 microscopic theory to calculate} the corresponding current--phase
relations and {presented numerical results
 illustrating their interference, under several physically motivated
 assumptions.}

{The resulting phenomenology is consistent with the
 Harvard group experiment}~\cite{Huang2022}. {In
 particular, our model can reproduce crossovers between noninversion
 and inversion as the bias voltage increases.} Split
quartets~\cite{Melin2020} {yield} noninversion or
inversion at zero bias depending on the sign of the corresponding
critical current. The electroflux
effect~\cite{Stoof-Nazarov1,Stoof-Nazarov2} {produces
 noninversion--inversion crossovers as a function of bias voltage.}

We also note that Eq.~(\ref{eq:Iqc-4}) features
{general $\theta$-shifted current--phase relations,
 arising from interference between supercurrent components that are
 odd and even in the superconducting phase variables.} Such
relations {resemble those in Andreev
 interferometers}~\cite{Wilhelm,Nakano,Zaitsev,Kadigrobov,Volkov,Stoof-Nazarov1,Stoof-Nazarov2,Yip,Belzig,Zaikin1,Zaikin2,Zaikin3,Zaikin4},
{where the Josephson and Aharonov--Bohm currents scale
 as $\sin\Phi$ and $\cos\Phi$, respectively, in the tunnel limit,
 with $\Phi$ being the magnetic flux through the loop. We have
 extended to MJJs several concepts recently developed for ballistic
 2D-metal-based Andreev interferometers}~\cite{Rashid2025,Melin2024}.

\section*{Acknowledgements}

R.M. and R.D. acknowledge the financial support from the SUPRADEVMAT
International Research Project between the French CNRS-Grenoble and
the German KIT-Karlsruhe. R.D. acknowledges the funding from the
Deutsche Forschungsgemeinschaft (DFG, German Research Foundation) --
467596333 and the support from the Helmholtz Association through
program NACIP. M.K. and A.S.R. acknowledge funding from the
Pennsylvania State University Materials Research Science and
Engineering Center supported by the US National Science Foundation
(DMR 2011839) and the US National Science Foundation (DMR 2415756).

\appendix

\section{Summary of known results}
\label{sec:known}
In this Appendix, we present a summary of some of the known results on
the quartet current-phase relations, starting with the three-terminal
quartets in subsection~\ref{sec:3Tq} and proceeding further with the
four-terminal split-quartets in subsection~\ref{sec:4Tsq}. The recent
results on the phase-AR in Andreev interferometers are summarized in
the forthcoming subsection~\ref{sec:phase-AR}.

\subsection{Three-terminal quartets}
\label{sec:3Tq}

In this subsection, we present the three-terminal quartet
current-phase relations, see Ref.~\onlinecite{Freyn2011} and the
Appendix~\ref{app:quartets}. In perturbation theory in the tunneling
amplitudes, the current-phase relations of the three-terminal quartets
originate from two Cooper pairs emanating from the condensate of the
grounded superconducting lead~$S_T$. Those split-Cooper pairs
reorganize by quantum-mechanically exchanging partners, and the
outgoing ones are transmitted into the condensates of $S_L$ and $S_R$
biased at the opposite voltages $V_{L,R}=\pm V$. This voltage biasing
condition produces the identical energies $4e V_T\equiv 2e(V_L+V_R)$
for the incoming and outgoing pairs of Cooper pairs. By the
time-energy uncertainty relation, this condition of energy
conservation produces a DC-supercurrent $I_q(V,\varphi_q)$ at the
finite bias voltage $V$, which is additionally sensitive to the only
static combination of the gauge-invariant phase variables
$\varphi_q=\varphi_L+\varphi_R-2\varphi_T$.

The current-phase relations of the three-terminal quartets in the
short-junction limit were obtained in Ref.~\onlinecite{Freyn2011} and
they are redemonstrated in Appendix~\ref{sec:Iq-short}, see
Eq.~(\ref{eq:Iq-short}):
\begin{eqnarray}
 \label{eq:quartets-short-1}
I_{q,3T,1}\left(\varphi_q,\Phi\right) &=& I_{q,3T,c,1}^{(odd)}
\sin\left(\varphi_q-\frac{4\pi\Phi}{\Phi_0}\right)\\
\label{eq:quartets-short-2}
I_{q,3T,2}\left(\varphi_q,\Phi\right)&=&I_{q,3T,c,2}^{(odd)}
\sin\varphi_q
,
\end{eqnarray}
where we used the gauge of Eqs.~(\ref{eq:gauge1})-(\ref{eq:gauge4}).
The three-terminal quartet current-phase relations are odd in the
superconducting phase variables, see Eq.~(\ref{eq:quartets-short-1})
and Eq.~(\ref{eq:quartets-short-2}).

\begin{figure}[htb]
\centerline{\includegraphics[width=\columnwidth]{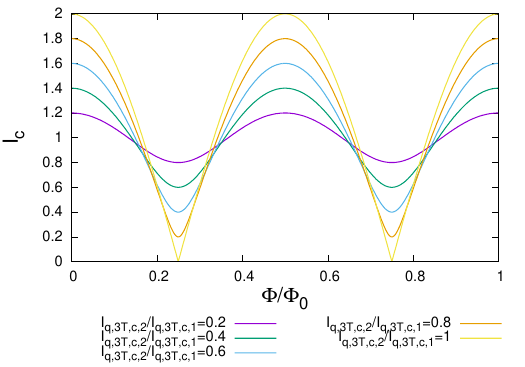}}
\caption{The first model for the interference $I_{q,3T,c}^{(1)}$, see
 Eq.~(\ref{eq:Iqc-1}). We evaluate the three-terminal quartet
 critical current for short-junctions, in the absence of
 nonequilibrium populations. The quartet critical current is plotted
 as a function of the reduced magnetic flux $\frac{\Phi}{\Phi_0}$.
 The values of the ratios
 $\frac{I_{q,3T,c,2}^{(odd)}}{I_{q,3T,c,1}^{(odd)}}$ between the
 corresponding critical currents of both three-terminal quartet
 junctions are shown on the figure. In the considered short-junction
 limit, those parameters are independent on the value of the bias
 voltage. This figure corresponds to the limiting case of a $4e$
 charge-Superconducting Quantum Interference Device (SQUID) with
 Josephson couplings that are symmetric or nonsymmetric.
\label{fig:interference1}
}
\end{figure}

{Now, we note that, in the device shown in
  Fig.~\ref{fig:device-4T}, the supercurrent transmitted into the loop
  is obtained as a sum over the amplitudes of the corresponding
  microscopic processes, as in a standard two-terminal Superconducting
  Quantum Interference Device (SQUID). Taking into account
  Eq.~(\ref{eq:quartets-short-1}) and Eq.~(\ref{eq:quartets-short-2})
  leads to the following oscillations of the critical current as a
  function of the magnetic flux $\Phi$:}
\begin{eqnarray}
 \label{eq:Iqc-1}
 I_{q,3T,c}^{(1)}\left(\Phi\right)&=&\mbox{Max}_{\varphi_q}\left[I_{q,3T,c,1}^{(odd)}
  \sin\left(\varphi_q-\frac{4\pi\Phi}{\Phi_0}\right)\right.\\&&\left. +
  I_{q,3T,c,2}^{(odd)} \sin\varphi_q\right] . \nonumber
\end{eqnarray}
Fig.~\ref{fig:interference1} shows the resulting {critical current}
$I_{q,3T,c}^{(1)} \left(\Phi\right)$ as a function of the reduced
magnetic flux $\frac{\Phi}{\Phi_0}$, for the ratios
$\frac{I_{q,3T,c,2}}{I_{q,3T,c,1}} = 0.2, \,0.4, \,0.6, \,0.8,
\,1$. Fig.~\ref{fig:interference1} reveals the expected
$\frac{1}{2}$-periodic interference as a function of
$\frac{\Phi}{\Phi_0}$, which is in agreement with the
$\frac{4\pi\Phi}{\Phi_0}$-sensitive oscillations in the above
Eq.~(\ref{eq:Iqc-1}). The interference of Eq.~(\ref{eq:Iqc-1}) is
analogous {to a SQUID with the charge $4e$.} This type
of interference was used in the experimental
Ref.~\onlinecite{Huang2022} to provide evidence for the $4e$-charge of
the Cooper quartets.

\subsection{Four-terminal split-quartets}
\label{sec:4Tsq}

In this subsection, we provide the expression of the split-quartet
current-phase relations, see Ref.~\onlinecite{Melin2020} and the
Appendix~\ref{app:quartets}. Those complementary four-terminal {\it
 split-quartets} exchange partners among the single Cooper pairs
coming from each of the grounded superconducting leads $S_R$ or
$S_B$. The resulting outgoing pairs are transmitted into $S_L$ and
$S_T$ biased at the opposite voltages $\pm V$. The four-terminal
split-quartet current-phase relations are redemonstrated in
Appendix~\ref{sec:Isq-short}, see Eq.~(\ref{eq:Isq-short}):
\begin{equation}
 \label{eq:sq-short}
 I_{sq}\left(\varphi_q,\Phi\right)=I_{sq,4T,c}^{(odd)}
 \sin\left(\varphi_q-\frac{2\pi\Phi}{\Phi_0}\right)
 ,
\end{equation}
where we used the gauge of Eqs.~(\ref{eq:gauge1})-(\ref{eq:gauge4}).

\begin{figure}[htb]
 \includegraphics[width=.48\textwidth]{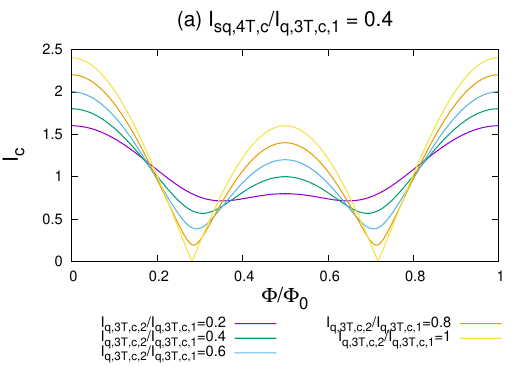}
 \includegraphics[width=.48\textwidth]{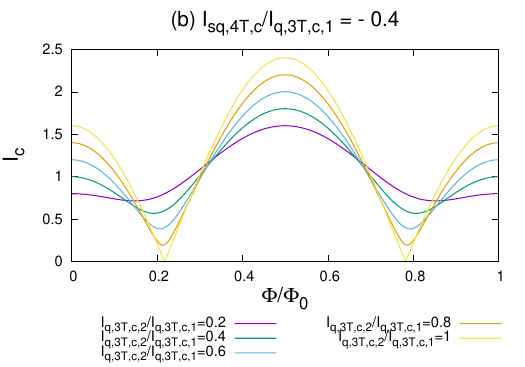}
\caption{The second model for the interference $I_{q,3T,c}^{(2)}$, see
 Eq.~(\ref{eq:Iqc-2}). We evaluate three- and four-terminal quartet
 and split-quartet critical current for short junctions, in the
 absence of nonequilibrium populations. The quartet critical current
 is plotted as a function of the reduced magnetic flux
 $\frac{\Phi}{\Phi_0}$. The values of the ratios
 $\frac{I_{q,3T,c,2}^{(odd)}}{I_{q,3T,c,1}^{(odd)}}$ and
 $\frac{I_{sq,4T,c}^{(odd)}}{I_{q,3T,c,1}^{(odd)}}$ between the
 corresponding critical currents are shown on the figure. In the
 considered short-junction limit, those parameters are independent on
 the value of the bias voltage. The second model produces
 $I_{q,3T,c}^{(2)}\left(\frac{\Phi}{\Phi_0}=\frac{1}{2}\right) <
 I_{q,3T,c}^{(2)}\left(\frac{\Phi}{\Phi_0}=0\right)$ or
 $I_{q,3T,c}^{(2)}\left(\frac{\Phi}{\Phi_0}=\frac{1}{2}\right)>I_{q,3T,c}^{(2)}\left(\frac{\Phi}{\Phi_0}=0\right)$,
 depending on the sign of the ratio
 $\frac{I_{sq,4T,c}^{(odd)}}{I_{q,3T,c,1}^{(odd)}}$.
 \label{fig:interference2}
}
\end{figure}

Now, we form an interference between the three-terminal quartets of
Eqs.~(\ref{eq:quartets-short-1})-(\ref{eq:quartets-short-2}) and the
four-terminal split quartets of Eq.~(\ref{eq:sq-short}), which all are
odd in the superconducting phase variables:
\begin{eqnarray}
 \label{eq:Iqc-2}
&& I_{q,3T,c}^{(2)}\left(\Phi\right)=\mbox{Max}_{\varphi_q}\left[I_{q,3T,c,1}^{(odd)}
  \sin\left(\varphi_q-\frac{4\pi\Phi}{\Phi_0}\right)\right.\\
  && + \left.
  I_{q,3T,c,2}^{(odd)} \sin\varphi_q +I_{sq,4T,c}^{(odd)}
  \sin\left(\varphi_q-\frac{2\pi\Phi}{\Phi_0}\right)
  \right] .
 \nonumber
\end{eqnarray}
The critical current in Fig.~\ref{fig:interference2} now breaks the
$\frac{\Phi}{\Phi_0}\rightarrow \frac{\Phi}{\Phi_0}+\frac{1}{2}$
symmetry, in agreement with the $\frac{2\pi\Phi}{\Phi_0}$-sensitivity
of the last term in the above Eq.~(\ref{eq:Iqc-2}). According to the
sign of the ratio $\frac{I_{sq,4T,c}^{(odd)}}{I_{q,3T,c}^{(odd)}}$
between the split-quartet and the quartet critical currents, we find
that $I_{q,3T,c}^{(2)}\left(\frac{\Phi}{\Phi_0}=\frac{1}{2}\right)$
can be smaller or larger than $I_{q,3T,c}^{(2)} \left(
\frac{\Phi}{\Phi_0} = 0 \right)$, corresponding the noninversion or to
the inversion respectively, see Ref.~\onlinecite{Melin2020}.

\begin{figure}[htb]
 \begin{minipage}{\columnwidth}\hspace*{-1.5cm}\includegraphics[width=1.2\textwidth]{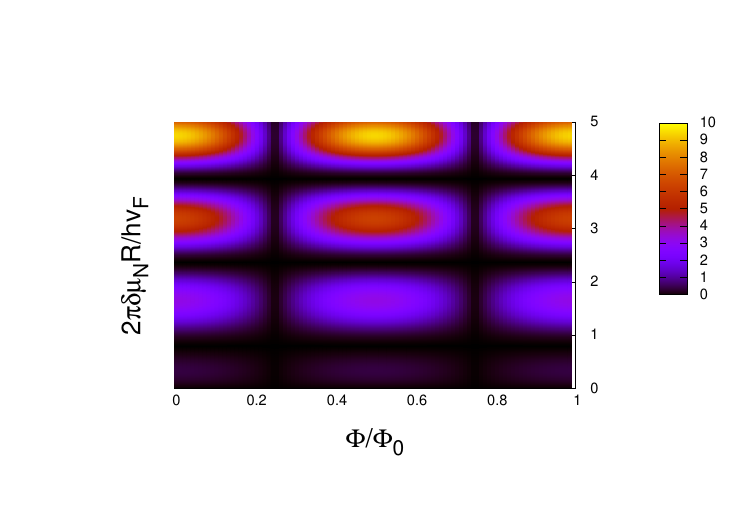}
  \end{minipage}
 \caption{The interference $I_{q,c}\left(\Phi,\delta\mu_N\right)$ in
  the critical current between the low-lying quartets emitted by the
  contacts $S_R$ and $S_B$, in the presence of coupling to the
  nonequilibrium quasiparticle populations in the central-$N$, see
  Eq.~(\ref{eq:Iqc-3}). The reduced magnetic flux
  $\frac{\Phi}{\Phi_0}$ is shown on the $x$-axis. The normalized
  shift $\delta \mu_N$ in the central-$N$ electrochemical potential
  is shown on the $y$-axis. The quartet critical current is shown in
  color in this parameter plane. Each junction individually taken is
  assumed to have the unity quartet critical current that is taken
  as the reference.
\label{fig:interference3}
}
\end{figure}

\subsection{Phase-Andreev reflection in an Andreev interferometer}
\label{sec:phase-AR}

In this subsection, we summarize recent results on a ballistic
Andreev interferometer, see
Refs.~\onlinecite{Melin2024,Rashid2025}. We note that a general
analogy holds with the two-fluid picture of superconductivity, where
the central-$N$ quasiparticle electrochemical potential $\delta
\mu_N\left(eV,\frac{\Phi}{\Phi_0}\right)$ parameterizes the
dissipative Andreev currents, being analogous to the normal
component. The superfluid component is analogous to the transport
processes of the nondissipative quartets and split-quartets. The
recently revisited theory of the Andreev interferometers
\cite{Melin2024,Rashid2025} relies on the coherent oscillatory
coupling between the analogous normal and superfluid components. This
coupling is operated by the coherent process of the phase-Andreev
reflection (phase-AR) which is here extended to the quartets in
multiterminal configurations. {In Andreev
 interferometers, the} phase-AR yields the dissipative Aharonov-Bohm
current that is proportional to the cosine of the flux $\Phi$ in the
limit of tunnel contacts.

\section{Calculation of the current}

In this Appendix, we provide a summary of the Keldysh algorithm by
which we calculate the current-phase relations in the vicinity of the
$x$-axis.

This model relies on low bias voltage on the superconducting
leads. This yields the adiabatic limit for the time-dependence of the
superconducting phase variables coexisting with nonequilibrium
quasiparticle populations. The above section~\ref{sec:FK} provides
arguments for why the Landau-Zener tunneling can be neglected.

The bare Keldysh Green's functions take the following form in such
quasi-equilibrium situation: $\hat{g}^{+,-}_{i,j}(\omega) =
n_F(\omega) \left[ \hat{g}^A_{i,j}(\omega) - \hat{g}^R_{i,j}(\omega)
 \right]$, where $n_F(\omega)$ is the electronic distribution
function of the central-$N$ as a function of the energy~$\omega$.

The spectral current $I_a(\omega)$ through the lead $S_a$ is expressed in
terms of the Keldysh Green's functions $\hat{G}^{+,-}_{\alpha,a}$ and
$\hat{G}^{+,-}_{a,\alpha}$:
\begin{equation}
 \label{eq:Ia}
 I_a(\omega)=\mbox{Nambu-trace}\left[\hat{\sigma_z}\left(\hat{\Sigma}_{a,\alpha}
  \hat{G}^{+,-}_{\alpha,a}-\hat{\Sigma}_{\alpha,a}
  \hat{G}^{+,-}_{a,\alpha}\right)\right] ,
\end{equation}
where ``Nambu-trace'' is a trace over the Nambu matrices, in the sense
of a summation over the ``1,1'' and ``2,2'' electron-electron and
hole-hole Nambu components. The notation $\hat{\sigma_z} =
\mbox{diag}(1, -1)$ stands for one of the Pauli matrices.

Using the expression
\cite{Caroli1,Caroli2,Cuevas}
\begin{equation}
 \hat{G}^{+,-}=\left(\hat{I}
 + \hat{G}^R \hat{\Sigma}\right)
 \hat{g}^{+,-} \left(\hat{I}
 + \hat{\Sigma} \hat{G}^A\right)
\end{equation}
of the Keldysh Green's function, we find the following decomposition
for $\hat{\Sigma}_{a,\alpha}\hat{G}^{+,-}_{\alpha,a}$, see the
expression of the current given by Eq.~(\ref{eq:Ia}):
\begin{equation}
 \label{eq:SplusN}
 \hat{\Sigma}_{a,\alpha}\hat{G}^{+,-}_{\alpha,a} =
  \left(\hat{\Sigma}_{a,\alpha}\hat{G}^{+,-}_{\alpha,a}\right)_S +
  \left(\hat{\Sigma}_{a,\alpha}\hat{G}^{+,-}_{\alpha,a}\right)_N
  ,
\end{equation}
where $\left( \hat{\Sigma}_{a,\alpha} \hat{G}^{+,-}_{\alpha,a}
\right)_S$ or $\left( \hat{\Sigma}_{a,\alpha} \hat{G}^{+,-}_{\alpha,a}
\right)_N$ involve the bare Keldysh Green's function in the
superconducting or normal leads. Perturbative expansions for $\left(
\hat{\Sigma}_{a,\alpha} \hat{G}^{+,-}_{\alpha,a} \right)_S$ and
$\left( \hat{\Sigma}_{a,\alpha} \hat{G}^{+,-}_{\alpha,a} \right)_N$
will be provided in Eqs.~(\ref{eq:greenS1})-(\ref{eq:greenS4}), and in
Eqs.~(\ref{eq:green1})-(\ref{eq:green4}) respectively. The bare
Keldysh Green's function in the infinite ballistic 2D-metal couples to
the density of states and populations, in a way that depends on the
shift $\delta \mu_N$ in the Fermi energy.

\section{Quartet current-phase relations}
\label{app:quartets}

Subsections~\ref{sec:Iq-short} and~\ref{sec:even-odd-long} provide the
microscopic calculations of the three-terminal quartets in the short-
and long-junction limits, respectively.
Subsection~\ref{sec:Isq-short} and~\ref{sec:sq-long} deal with the
four-terminal split-quartets in the short- and long-junction limits,
respectively.

\begin{figure*}[htb]
\centerline{\includegraphics[width=.3\textwidth]{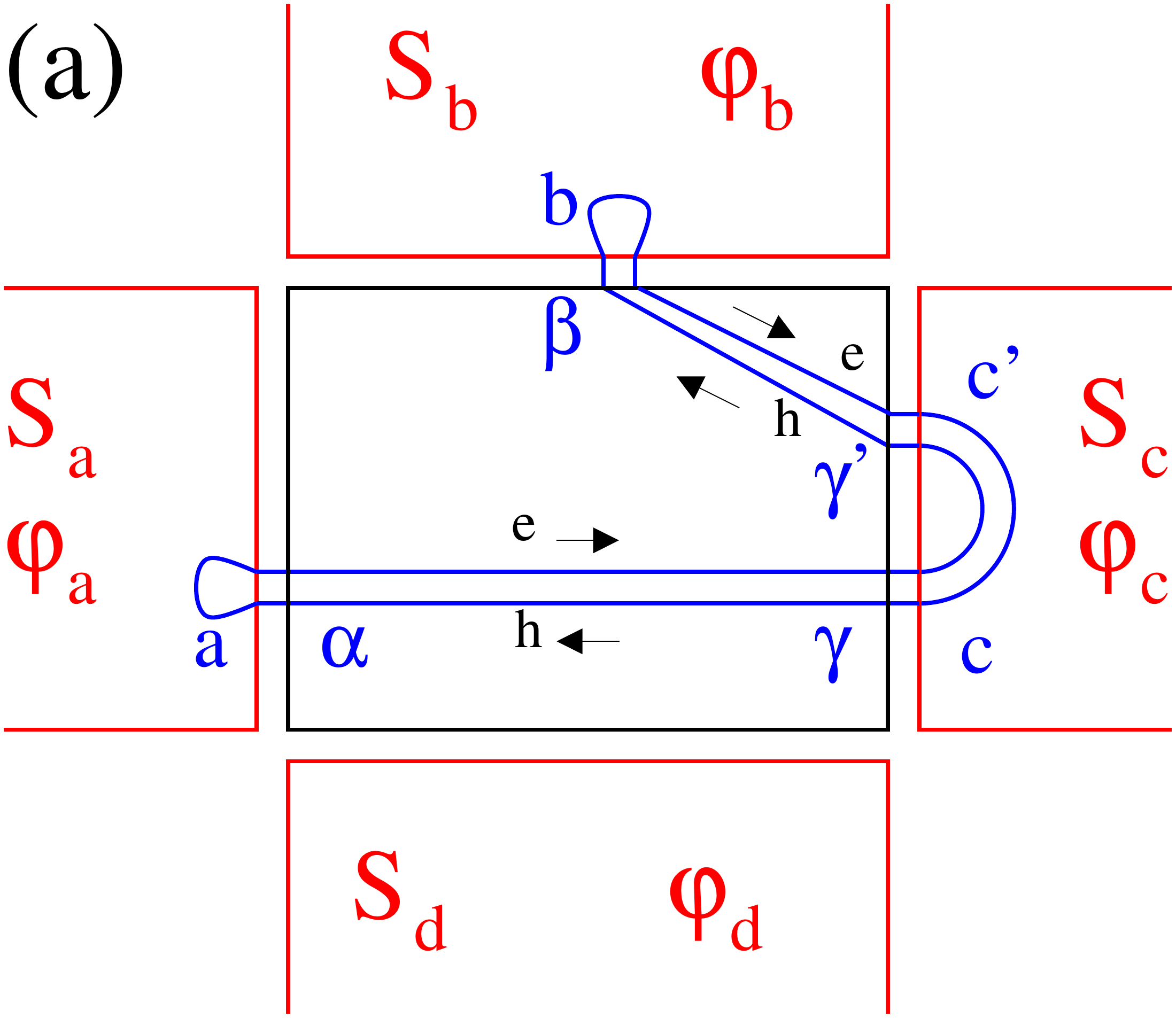} \hspace{1cm}
 \includegraphics[width=.3\textwidth]{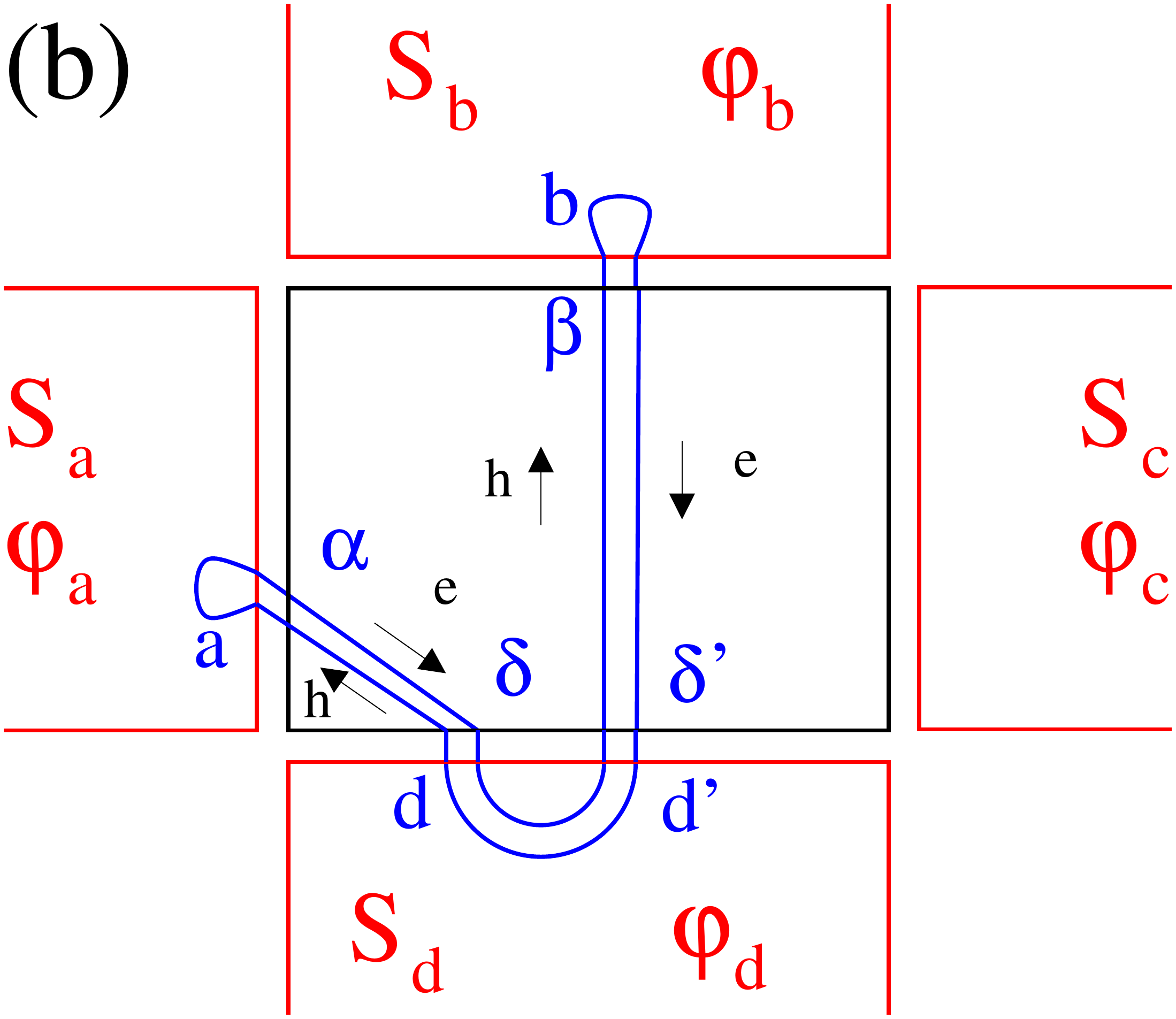}}
\caption{The three-terminal quartets. Two pairs are emitted from
 $S_a$ and $S_b$, and absorbed into $S_c$ (a) or $S_d$ (b).
\label{fig:3T-quartets}
}
\end{figure*}

\subsection{Odd-in-phase/even-in-voltage
 three-terminal quartets}

\label{sec:Iq-short}

We previously decomposed Eq.~(\ref{eq:SplusN}) for
$\hat{\Sigma}_{a,\alpha}\hat{G}^{+,-}_{\alpha,a}$ into both terms
$\left(\hat{\Sigma}_{a,\alpha}\hat{G}^{+,-}_{\alpha,a}\right)_S$ and
$\left(\hat{\Sigma}_{a,\alpha}\hat{G}^{+,-}_{\alpha,a}\right)_N$
having the bare Keldysh Green's function in the superconducting or
normal leads, respectively. In this subsection, we now evaluate
$\left(\hat{\Sigma}_{a,\alpha}\hat{G}^{+,-}_{\alpha,a}\right)_S$ in
perturbation in the hopping amplitudes, and we will confirm the form
of the quartet critical current in the short-junction limit, in
agreement with Ref.~\onlinecite{Freyn2011} and with the above
Eq.~(\ref{eq:Iqc-1}).

Concerning perturbation theory in the hopping amplitudes, we expand
$\left(\hat{\Sigma}_{a,\alpha}\hat{G}^{+,-}_{\alpha,a}\right)_S$ in
perturbation in the hopping amplitudes, to the lowest order that
produces the emergence of the quartets, i.e. to the order $\Sigma_a^2
\Sigma_b^2 \Sigma_c^2 \Sigma_{c'}^2$:
\begin{widetext}
\begin{eqnarray}
 \label{eq:greenS1}
\left(\hat{\Sigma}_{a,\alpha}^{1,1} \hat{G}^{+,-,1,1}_{\alpha,a}\right)_S
 &=&
 \hat{\Sigma}_{a,\alpha}^{1,1}
 \hat{g}^{1,1,R}_{\alpha,\gamma}
 \hat{\Sigma}_{\gamma,c}^{1,1}
 \hat{g}_{c,c'}^{1,2,+,-}
 \hat{\Sigma}_{c',\gamma'}^{2,2}
 \hat{g}_{\gamma',\beta}^{2,2,A}
 \hat{\Sigma}_{\beta,b}^{2,2}
 \hat{g}^{2,1,A}_{b,b}
 \hat{\Sigma}_{b,\beta}^{1,1}
 \hat{g}^{1,1,A}_{\beta,\gamma'}
 \hat{\Sigma}_{\gamma',c'}^{1,1}
 \hat{g}_{c',c}^{1,2,A}
 \hat{\Sigma}_{c,\gamma}^{2,2}
 \hat{g}_{\gamma,\alpha}^{2,2,A}
 \hat{\Sigma}_{\alpha,a}^{2,2}
 \hat{g}_{a,a}^{2,1,A}\\
 &+&
 \hat{\Sigma}_{a,\alpha}^{1,1}
 \hat{g}^{1,1,R}_{\alpha,\gamma}
 \hat{\Sigma}_{\gamma,c}^{1,1}
 \hat{g}_{c,c'}^{1,2,R}
 \hat{\Sigma}_{c',\gamma'}^{2,2}
 \hat{g}_{\gamma',\beta}^{2,2,R}
 \hat{\Sigma}_{\beta,b}^{2,2}
 \hat{g}^{2,1,+,-}_{b,b}
 \hat{\Sigma}_{b,\beta}^{1,1}
 \hat{g}^{1,1,A}_{\beta,\gamma'}
 \hat{\Sigma}_{\gamma',c'}^{1,1}
 \hat{g}_{c',c}^{1,2,A}
 \hat{\Sigma}_{c,\gamma}^{2,2}
 \hat{g}_{\gamma,\alpha}^{2,2,A}
 \hat{\Sigma}_{\alpha,a}^{2,2}
 \hat{g}_{a,a}^{2,1,A}\\
&+&
 \hat{\Sigma}_{a,\alpha}^{1,1}
 \hat{g}^{1,1,R}_{\alpha,\gamma}
 \hat{\Sigma}_{\gamma,c}^{1,1}
 \hat{g}_{c,c'}^{1,2,R}
 \hat{\Sigma}_{c',\gamma'}^{2,2}
 \hat{g}_{\gamma',\beta}^{2,2,R}
 \hat{\Sigma}_{\beta,b}^{2,2}
 \hat{g}^{2,1,R}_{b,b}
 \hat{\Sigma}_{b,\beta}^{1,1}
 \hat{g}^{1,1,R}_{\beta,\gamma'}
 \hat{\Sigma}_{\gamma',c'}^{1,1}
 \hat{g}_{c',c}^{1,2,+,-}
 \hat{\Sigma}_{c,\gamma}^{2,2}
 \hat{g}_{\gamma,\alpha}^{2,2,A}
 \hat{\Sigma}_{\alpha,a}^{2,2}
 \hat{g}_{a,a}^{2,1,A}\\
&+&
 \hat{\Sigma}_{a,\alpha}^{1,1}
 \hat{g}^{1,1,R}_{\alpha,\gamma}
 \hat{\Sigma}_{\gamma,c}^{1,1}
 \hat{g}_{c,c'}^{1,2,R}
 \hat{\Sigma}_{c',\gamma'}^{2,2}
 \hat{g}_{\gamma',\beta}^{2,2,R}
 \hat{\Sigma}_{\beta,b}^{2,2}
 \hat{g}^{2,1,R}_{b,b}
 \hat{\Sigma}_{b,\beta}^{1,1}
 \hat{g}^{1,1,R}_{\beta,\gamma'}
 \hat{\Sigma}_{\gamma',c'}^{1,1}
 \hat{g}_{c',c}^{1,2,R}
 \hat{\Sigma}_{c,\gamma}^{2,2}
 \hat{g}_{\gamma,\alpha}^{2,2,R}
 \hat{\Sigma}_{\alpha,a}^{2,2}
 \hat{g}_{a,a}^{2,1,+,-}
 .
 \label{eq:greenS4}
\end{eqnarray}
The bare Keldysh Green's functions $\hat{g}^{+,-}_{i,j}(\omega)$ are
nonvanishingly small in the presence of a finite superconducting gap,
and Eqs.~(\ref{eq:greenS1})-(\ref{eq:greenS4}) are expanded as
follows:
\begin{eqnarray}
 \label{eq:greenS-simplified-debut}
\left(\hat{\Sigma}_{a,\alpha}^{1,1} \hat{G}^{+,-,1,1}_{\alpha,a}\right)_S
 &=& n_F(\omega)
 \hat{\Sigma}_{a,\alpha}^{1,1}
 \hat{g}^{1,1,R}_{\alpha,\gamma}
 \hat{\Sigma}_{\gamma,c}^{1,1}
 \hat{g}_{c,c'}^{1,2,A}
 \hat{\Sigma}_{c',\gamma'}^{2,2}
 \hat{g}_{\gamma',\beta}^{2,2,A}
 \hat{\Sigma}_{\beta,b}^{2,2}
 \hat{g}^{2,1,A}_{b,b}
 \hat{\Sigma}_{b,\beta}^{1,1}
 \hat{g}^{1,1,A}_{\beta,\gamma'}
 \hat{\Sigma}_{\gamma',c'}^{1,1}
 \hat{g}_{c',c}^{1,2,A}
 \hat{\Sigma}_{c,\gamma}^{2,2}
 \hat{g}_{\gamma,\alpha}^{2,2,A}
 \hat{\Sigma}_{\alpha,a}^{2,2}
 \hat{g}_{a,a}^{2,1,A}\\
 \label{eq:TOTO3}
 &-&
  n_F(\omega)
 \hat{\Sigma}_{a,\alpha}^{1,1}
 \hat{g}^{1,1,R}_{\alpha,\gamma}
 \hat{\Sigma}_{\gamma,c}^{1,1}
 \hat{g}_{c,c'}^{1,2,R}
 \hat{\Sigma}_{c',\gamma'}^{2,2}
 \hat{g}_{\gamma',\beta}^{2,2,A}
 \hat{\Sigma}_{\beta,b}^{2,2}
 \hat{g}^{2,1,A}_{b,b}
 \hat{\Sigma}_{b,\beta}^{1,1}
 \hat{g}^{1,1,A}_{\beta,\gamma'}
 \hat{\Sigma}_{\gamma',c'}^{1,1}
 \hat{g}_{c',c}^{1,2,A}
 \hat{\Sigma}_{c,\gamma}^{2,2}
 \hat{g}_{\gamma,\alpha}^{2,2,A}
 \hat{\Sigma}_{\alpha,a}^{2,2}
 \hat{g}_{a,a}^{2,1,A}
 \\
 \label{eq:TOTO4}
 &+& n_F(\omega)
 \hat{\Sigma}_{a,\alpha}^{1,1}
 \hat{g}^{1,1,R}_{\alpha,\gamma}
 \hat{\Sigma}_{\gamma,c}^{1,1}
 \hat{g}_{c,c'}^{1,2,R}
 \hat{\Sigma}_{c',\gamma'}^{2,2}
 \hat{g}_{\gamma',\beta}^{2,2,R}
 \hat{\Sigma}_{\beta,b}^{2,2}
 \hat{g}^{2,1,A}_{b,b}
 \hat{\Sigma}_{b,\beta}^{1,1}
 \hat{g}^{1,1,A}_{\beta,\gamma'}
 \hat{\Sigma}_{\gamma',c'}^{1,1}
 \hat{g}_{c',c}^{1,2,A}
 \hat{\Sigma}_{c,\gamma}^{2,2}
 \hat{g}_{\gamma,\alpha}^{2,2,A}
 \hat{\Sigma}_{\alpha,a}^{2,2}
 \hat{g}_{a,a}^{2,1,A}\\
 \label{eq:TOTO1}
 &-&
  n_F(\omega)
 \hat{\Sigma}_{a,\alpha}^{1,1}
 \hat{g}^{1,1,R}_{\alpha,\gamma}
 \hat{\Sigma}_{\gamma,c}^{1,1}
 \hat{g}_{c,c'}^{1,2,R}
 \hat{\Sigma}_{c',\gamma'}^{2,2}
 \hat{g}_{\gamma',\beta}^{2,2,R}
 \hat{\Sigma}_{\beta,b}^{2,2}
 \hat{g}^{2,1,R}_{b,b}
 \hat{\Sigma}_{b,\beta}^{1,1}
 \hat{g}^{1,1,A}_{\beta,\gamma'}
 \hat{\Sigma}_{\gamma',c'}^{1,1}
 \hat{g}_{c',c}^{1,2,A}
 \hat{\Sigma}_{c,\gamma}^{2,2}
 \hat{g}_{\gamma,\alpha}^{2,2,A}
 \hat{\Sigma}_{\alpha,a}^{2,2}
 \hat{g}_{a,a}^{2,1,A}\\
 \label{eq:TOTO2}
&+& n_F(\omega)
 \hat{\Sigma}_{a,\alpha}^{1,1}
 \hat{g}^{1,1,R}_{\alpha,\gamma}
 \hat{\Sigma}_{\gamma,c}^{1,1}
 \hat{g}_{c,c'}^{1,2,R}
 \hat{\Sigma}_{c',\gamma'}^{2,2}
 \hat{g}_{\gamma',\beta}^{2,2,R}
 \hat{\Sigma}_{\beta,b}^{2,2}
 \hat{g}^{2,1,R}_{b,b}
 \hat{\Sigma}_{b,\beta}^{1,1}
 \hat{g}^{1,1,R}_{\beta,\gamma'}
 \hat{\Sigma}_{\gamma',c'}^{1,1}
 \hat{g}_{c',c}^{1,2,A}
 \hat{\Sigma}_{c,\gamma}^{2,2}
 \hat{g}_{\gamma,\alpha}^{2,2,A}
 \hat{\Sigma}_{\alpha,a}^{2,2}
 \hat{g}_{a,a}^{2,1,A}
 \\
 \label{eq:TOTO5}
&-& n_F(\omega)
 \hat{\Sigma}_{a,\alpha}^{1,1}
 \hat{g}^{1,1,R}_{\alpha,\gamma}
 \hat{\Sigma}_{\gamma,c}^{1,1}
 \hat{g}_{c,c'}^{1,2,R}
 \hat{\Sigma}_{c',\gamma'}^{2,2}
 \hat{g}_{\gamma',\beta}^{2,2,R}
 \hat{\Sigma}_{\beta,b}^{2,2}
 \hat{g}^{2,1,R}_{b,b}
 \hat{\Sigma}_{b,\beta}^{1,1}
 \hat{g}^{1,1,R}_{\beta,\gamma'}
 \hat{\Sigma}_{\gamma',c'}^{1,1}
 \hat{g}_{c',c}^{1,2,R}
 \hat{\Sigma}_{c,\gamma}^{2,2}
 \hat{g}_{\gamma,\alpha}^{2,2,A}
 \hat{\Sigma}_{\alpha,a}^{2,2}
 \hat{g}_{a,a}^{2,1,A}
 \\
 \label{eq:TOTO6}
&+& n_F(\omega)
 \hat{\Sigma}_{a,\alpha}^{1,1}
 \hat{g}^{1,1,R}_{\alpha,\gamma}
 \hat{\Sigma}_{\gamma,c}^{1,1}
 \hat{g}_{c,c'}^{1,2,R}
 \hat{\Sigma}_{c',\gamma'}^{2,2}
 \hat{g}_{\gamma',\beta}^{2,2,R}
 \hat{\Sigma}_{\beta,b}^{2,2}
 \hat{g}^{2,1,R}_{b,b}
 \hat{\Sigma}_{b,\beta}^{1,1}
 \hat{g}^{1,1,R}_{\beta,\gamma'}
 \hat{\Sigma}_{\gamma',c'}^{1,1}
 \hat{g}_{c',c}^{1,2,R}
 \hat{\Sigma}_{c,\gamma}^{2,2}
 \hat{g}_{\gamma,\alpha}^{2,2,R}
 \hat{\Sigma}_{\alpha,a}^{2,2}
 \hat{g}_{a,a}^{2,1,A}\\
 &-&
 n_F(\omega)
 \hat{\Sigma}_{a,\alpha}^{1,1}
 \hat{g}^{1,1,R}_{\alpha,\gamma}
 \hat{\Sigma}_{\gamma,c}^{1,1}
 \hat{g}_{c,c'}^{1,2,R}
 \hat{\Sigma}_{c',\gamma'}^{2,2}
 \hat{g}_{\gamma',\beta}^{2,2,R}
 \hat{\Sigma}_{\beta,b}^{2,2}
 \hat{g}^{2,1,R}_{b,b}
 \hat{\Sigma}_{b,\beta}^{1,1}
 \hat{g}^{1,1,R}_{\beta,\gamma'}
 \hat{\Sigma}_{\gamma',c'}^{1,1}
 \hat{g}_{c',c}^{1,2,R}
 \hat{\Sigma}_{c,\gamma}^{2,2}
 \hat{g}_{\gamma,\alpha}^{2,2,R}
 \hat{\Sigma}_{\alpha,a}^{2,2}
 \hat{g}_{a,a}^{2,1,R}
 . \label{eq:greenS-simplified-fin}
\end{eqnarray}

Using the results of the forthcoming
{Appendix~\ref{app:averaging},} the above
Eqs.~(\ref{eq:greenS-simplified-debut})-(\ref{eq:greenS-simplified-fin})
are found to take the following values:
(\ref{eq:greenS-simplified-debut})$=-X \left(Y^A\right)^4$,
(\ref{eq:TOTO3})$=$(\ref{eq:TOTO4})$=X \left(Y^A\right)^3 Y^R$,
(\ref{eq:TOTO1})$=$(\ref{eq:TOTO2})$=-X\left(Y^A\right)^2
\left(Y^R\right)^2$,
(\ref{eq:TOTO5})$=$(\ref{eq:TOTO6})$=X\left(Y^R\right)^3 Y^A$ and
(\ref{eq:greenS-simplified-fin})$=-X\left(Y^R\right)^4$, where we used
the following short notations:
\begin{eqnarray}
 \label{eq:X}
 X&=&\frac{\Sigma_a^2 \Sigma_b^2 \Sigma_c^2 \Sigma_{c'}^2
  n_F(\omega)}{4 W^8 (k_F R_{\alpha,\gamma})(k_F R_{\beta,\gamma'})}
 e^{i(2\varphi_c-\varphi_a-\varphi_b)} \cos\left(\frac{2\omega
  R_{\alpha,\gamma}}{\hbar v_F}\right) \cos\left(\frac{2\omega
  R_{\beta,\gamma'}}{\hbar v_F}\right)\\
\label{eq:YA}
Y^A&=&\left(\frac{\Delta^2}{\Delta^2-(\omega-i\eta)^2}\right)^{\frac{1}{2}}\\
\label{eq:YR}
 Y^R&=&\left(\frac{\Delta^2}{\Delta^2-(\omega+i\eta)^2}\right)^{\frac{1}{2}}
 .
\end{eqnarray}
We deduce
\begin{equation}
 \overline{(\mbox{\ref{eq:greenS-simplified-debut}})+...+(\mbox{\ref{eq:greenS-simplified-fin})}}
 =-X\left[\left(Y^A\right)^2+\left(Y^R\right)^2\right]\left[Y^A-Y^R\right]^2
 .
\end{equation}
Subtracting the ``2,2'' component leads to
\begin{eqnarray}
 \label{eq:diff1}
 &&
 \overline{\left(\hat{\Sigma}_{a,\alpha}^{1,1} \hat{G}^{+,-,1,1}_{\alpha,a}\right)_S}
-
\overline{\left(\hat{\Sigma}_{a,\alpha}^{2,2} \hat{G}^{+,-,2,2}_{\alpha,a}\right)_S}\\
\nonumber
&=&
i \frac{\Sigma_a^2 \Sigma_b^2 \Sigma_c^2 \Sigma_{c'}^2
  n_F(\omega)}{2 W^8 (k_F R_{\alpha,\gamma})(k_F R_{\beta,\gamma'})}
\sin\left(2\varphi_c-\varphi_a-\varphi_b\right)
 \cos\left(\frac{2\omega R_{\alpha,\gamma}}{\hbar v_F}\right)
 \cos\left(\frac{2\omega R_{\beta,\gamma'}}{\hbar
  v_F}\right)\times\\
&& \left[\left(\frac{\Delta^2}{\Delta^2-(\omega-i\eta)^2}\right)
 +
 \left(\frac{\Delta^2}{\Delta^2-(\omega+i\eta)^2}\right)\right]\times
 \left[\left(\frac{\Delta^2}{\Delta^2-(\omega-i\eta)^2}\right)^{\frac{1}{2}}
  -
  \left(\frac{\Delta^2}{\Delta^2-(\omega+i\eta)^2}\right)^{\frac{1}{2}}
  \right]^2
 ,
 \nonumber
\end{eqnarray}
\end{widetext}
and the quartet current is given by
\begin{eqnarray}
 \label{eq:diff2}
 &&I_a=\\&&2 \int \mbox{Re}\left[\overline{\left(\hat{\Sigma}_{a,\alpha}^{1,1}
  \hat{G}^{+,-,1,1}_{\alpha,a}\right)_S} -
  \overline{\left(\hat{\Sigma}_{a,\alpha}^{2,2}
   \hat{G}^{+,-,2,2}_{\alpha,a}\right)_S} \right] d\omega
 \nonumber
 ,
\end{eqnarray}
where Eq.~(\ref{eq:diff1}) should be inserted into
Eq.~(\ref{eq:diff2}). We deduce that the quartet current takes the
following form at the lowest order in the tunneling amplitudes and in
the short-junction limit $\frac{2\Delta R_{\alpha,\gamma}}{\hbar v_F}\alt 1$
and $\frac{2\Delta R_{\beta,\gamma'}}{\hbar v_F}\alt 1$:
\begin{equation}
 \label{eq:Iq-short}
I_q^{(short)}=I_{q,3T,c}^{(short)} \sin\left(2\varphi_c-\varphi_a-\varphi_b\right)
,
\end{equation}
where $I_{q,3T,c}^{(short)}$ is a critical current that is independent
on the bias voltage $V$. {Eq.~(\ref{eq:Iq-short}) is
 used in the main text, see
 Eqs.~(\ref{eq:quartets-short-1})-(\ref{eq:quartets-short-2}).}

\subsection{Even-in-phase/odd-in-voltage three-terminal quartets}
\label{sec:even-odd-long}
In this subsection, we consider the long-junction limit
\cite{Kulik,Ishii,Bagwell}, which hosts a discrete spectrum of ABS, at
all energy scales with respect to the superconducting gap. We
specifically calculate the corresponding quartet current as a function
of both the quartet phase and the bias voltage.

We focus on small energies $\omega\ll \Delta$ compared to the
superconducting gap $\Delta$, and thus, we implement the large-gap
approximation to evaluate the second term
$\left(\hat{\Sigma}_{a,\alpha}\hat{G}^{+,-}_{\alpha,a}\right)_N$ in
Eq.~(\ref{eq:SplusN}):
\begin{widetext}
\begin{eqnarray}
 \label{eq:green1}
 \left(\hat{\Sigma}_{a,\alpha}^{1,1} \hat{G}^{+,-,1,1}_{\alpha,a}\right)_N
 &=&
 \hat{\Sigma}_{a,\alpha}^{1,1}
 \hat{g}^{1,1,+,-}_{\alpha,\gamma}
 \hat{\Sigma}_{\gamma,c}^{1,1}
 \hat{g}_{c,c'}^{1,2,A}
 \hat{\Sigma}_{c',\gamma'}^{2,2}
 \hat{g}_{\gamma',\beta}^{2,2,A}
 \hat{\Sigma}_{\beta,b}^{2,2}
 \hat{g}^{2,1,A}_{b,b}
 \hat{\Sigma}_{b,\beta}^{1,1}
 \hat{g}^{1,1,A}_{\beta,\gamma'}
 \hat{\Sigma}_{\gamma',c'}^{1,1}
 \hat{g}_{c',c}^{1,2,A}
 \hat{\Sigma}_{c,\gamma}^{2,2}
 \hat{g}_{\gamma,\alpha}^{2,2,A}
 \hat{\Sigma}_{\alpha,a}^{2,2}
 \hat{g}_{a,a}^{2,1,A}\\
 \label{eq:green2}
 &+&
 \hat{\Sigma}_{a,\alpha}^{1,1}
 \hat{g}^{1,1,R}_{\alpha,\gamma}
 \hat{\Sigma}_{\gamma,c}^{1,1}
 \hat{g}_{c,c'}^{1,2,R}
 \hat{\Sigma}_{c',\gamma'}^{2,2}
 \hat{g}_{\gamma',\beta}^{2,2,+,-}
 \hat{\Sigma}_{\beta,b}^{2,2}
 \hat{g}^{2,1,A}_{b,b}
 \hat{\Sigma}_{b,\beta}^{1,1}
 \hat{g}^{1,1,A}_{\beta,\gamma'}
 \hat{\Sigma}_{\gamma',c'}^{1,1}
 \hat{g}_{c',c}^{1,2,A}
 \hat{\Sigma}_{c,\gamma}^{2,2}
 \hat{g}_{\gamma,\alpha}^{2,2,A}
 \hat{\Sigma}_{\alpha,a}^{2,2}
 \hat{g}_{a,a}^{2,1,A}\\
 \label{eq:green3}
 &+&
 \hat{\Sigma}_{a,\alpha}^{1,1}
 \hat{g}^{1,1,R}_{\alpha,\gamma}
 \hat{\Sigma}_{\gamma,c}^{1,1}
 \hat{g}_{c,c'}^{1,2,R}
 \hat{\Sigma}_{c',\gamma'}^{2,2}
 \hat{g}_{\gamma',\beta}^{2,2,R}
 \hat{\Sigma}_{\beta,b}^{2,2}
 \hat{g}^{2,1,R}_{b,b}
 \hat{\Sigma}_{b,\beta}^{1,1}
 \hat{g}^{1,1,+,-}_{\beta,\gamma'}
 \hat{\Sigma}_{\gamma',c'}^{1,1}
 \hat{g}_{c',c}^{1,2,A}
 \hat{\Sigma}_{c,\gamma}^{2,2}
 \hat{g}_{\gamma,\alpha}^{2,2,A}
 \hat{\Sigma}_{\alpha,a}^{2,2}
 \hat{g}_{a,a}^{2,1,A}\\
 &+&
 \hat{\Sigma}_{a,\alpha}^{1,1}
 \hat{g}^{1,1,R}_{\alpha,\gamma}
 \hat{\Sigma}_{\gamma,c}^{1,1}
 \hat{g}_{c,c'}^{1,2,R}
 \hat{\Sigma}_{c',\gamma'}^{2,2}
 \hat{g}_{\gamma',\beta}^{2,2,R}
 \hat{\Sigma}_{\beta,b}^{2,2}
 \hat{g}^{2,1,R}_{b,b}
 \hat{\Sigma}_{b,\beta}^{1,1}
 \hat{g}^{1,1,R}_{\beta,\gamma'}
 \hat{\Sigma}_{\gamma',c'}^{1,1}
 \hat{g}_{c',c}^{1,2,R}
 \hat{\Sigma}_{c,\gamma}^{2,2}
 \hat{g}_{\gamma,\alpha}^{2,2,+,-}
 \hat{\Sigma}_{\alpha,a}^{2,2}
 \hat{g}_{a,a}^{2,1,A}
 \label{eq:green4}
 .
\end{eqnarray}
\end{widetext}

We now evaluate the four terms given by
Eqs.~(\ref{eq:green1})-(\ref{eq:green4}) and average out the fast
oscillations at the smallest length scale of the Fermi wave-length,
see the forthcoming Appendix~\ref{app:averaging}. Then, combining
Eq.~(\ref{eq:Ia}) to Eqs.~(\ref{eq:green1})-(\ref{eq:green4}) leads to
the following expression of the spectral current:
\begin{eqnarray}
 \nonumber
 \overline{I_a}(\omega)&=&\frac{2 \Sigma_a^2 \Sigma_b^2 \Sigma_c^4}{W^8
  (k_F R_{\alpha,\gamma}) (k_F R_{\beta,\gamma'})}
 \cos\left(\frac{2\delta \mu_N R_{\alpha,\gamma}}{\hbar v_F}\right)\times\\&&
 \nonumber
 \cos\left(\frac{2\delta \mu_N R_{\beta,\gamma'}}{\hbar v_F}\right)
 \cos\left(\varphi_a+\varphi_b-2\varphi_c\right)\times\\&&
 \left[n_F(\omega-\delta \mu_N)-n_F(\omega+\delta\mu_N)\right]
 ,
 \label{eq:Ia-phiQ}
\end{eqnarray}
where the overline denotes averaging at the scale of the Fermi
wave-length. Eq.~(\ref{eq:Ia-phiQ}) contributes to the mesoscopic
oscillations of the quartet current, as a function of the
2D-central-$N$ shift in the Fermi energy $\delta\mu_N$, and as a
function of the quartet phase.

To summarize, we find that the quartet current takes the following
form at the lowest order in the hopping amplitudes and in the
long-junction limit:
\begin{eqnarray}
 \nonumber
 I_q^{(long)}&=& I_{q,3T,c}^{(even)} \delta\mu_N \cos\left(\frac{2\delta
  \mu_N R_{\alpha,\gamma}}{\hbar v_F}\right) \cos\left(\frac{2\delta
 \mu_N R_{\beta,\gamma'}}{\hbar v_F}\right)\\&&
 \cos\left(\varphi_a+\varphi_b-2\varphi_c\right)
 ,
 \label{eq:Iq-long}
\end{eqnarray}
where we additionally assumed that $\delta \mu_N$ is small compared to
the superconducting gap. {Eq.~(\ref{eq:Iq-long}) in
 used in the main text, see Eq.~(\ref{eq:Iqceff}).}

\begin{figure*}[htb]
\centerline{\includegraphics[width=.3\textwidth]{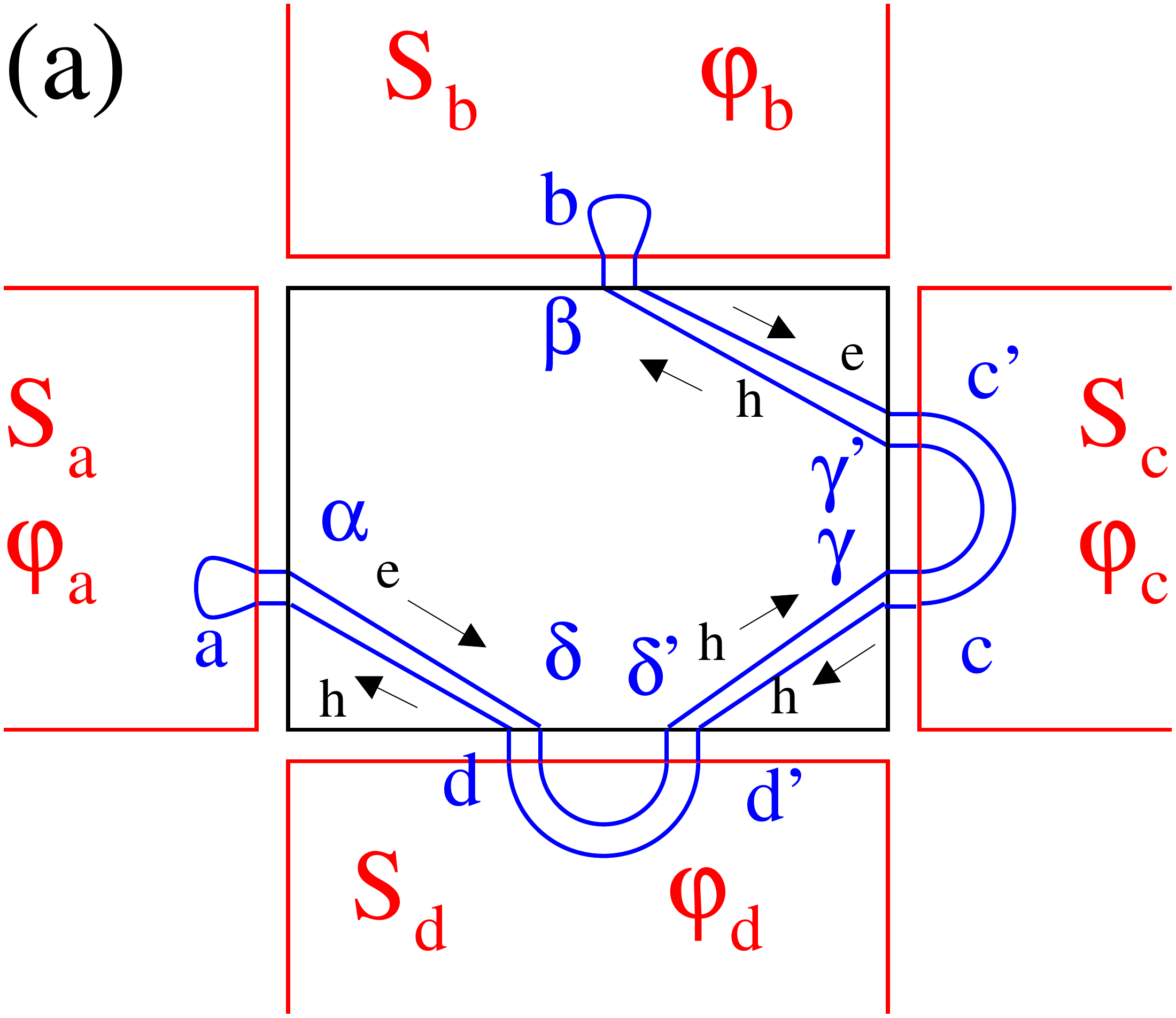} \hspace{1cm}
 \includegraphics[width=.3\textwidth]{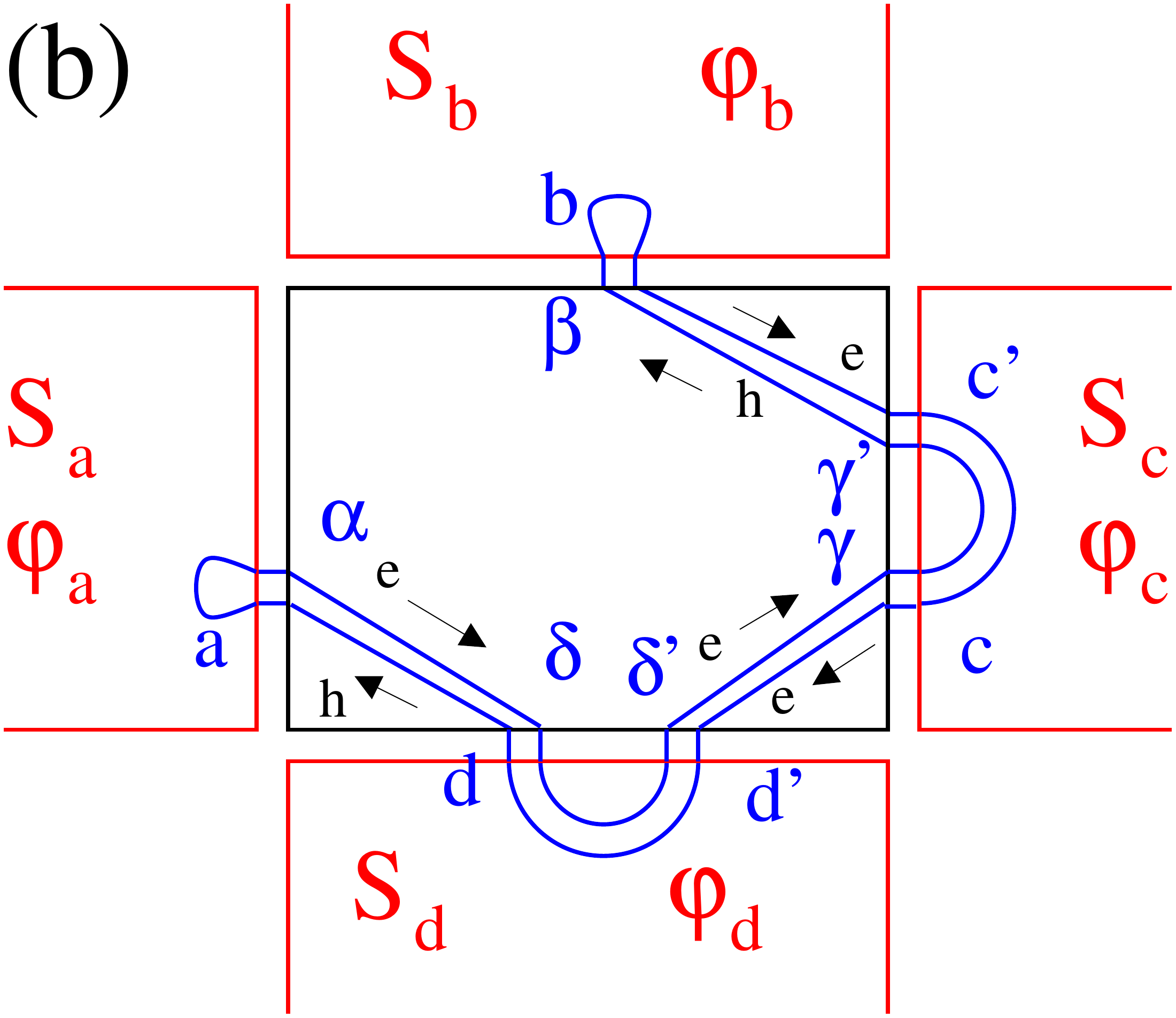}}
\caption{The four-terminal split-quartets. Two pairs are emitted from
 $S_a$ and $S_b$, they recombine and are separately absorbed into
 $S_c$ and $S_d$. Two complementary processes of four-terminal
 split-quartets appear if the superconducting leads $S_c$ and $S_d$
 are interchanged.
\label{fig:4T-quartets}
}
\end{figure*}

\subsection{Odd-in-phase/even-in-voltage four-terminal split-quartets}
\label{sec:Isq-short}

In this subsection, we go back to the short-junction limit and
consider the four-terminal device in Fig.~\ref{fig:device-4T},
sustaining the process of the split-quartets. Namely, the two incoming
pairs are taken from the grounded $S_c$ and $S_d$, they exchange
partners and are transmitted as the two outgoing pairs into the
superconducting leads $S_a$ and $S_b$ biased at the voltages $\pm
V$. In this subsection of Appendix~\ref{app:quartets}, we first
recalculate the split-quartet critical current in the short-junction
limit.

The expression of the four-terminal split-quartet diagrams in
Fig.~\ref{fig:4T-quartets} are the following:
\begin{equation}
 \label{eq:Sigma-G}
 \left(\hat{\Sigma}_{a,\alpha}\hat{G}_{\alpha,a}\right)_N
 =
  \left(\hat{\Sigma}_{a,\alpha}\hat{G}_{\alpha,a}\right)_N^{(A)}
  +
  \left(\hat{\Sigma}_{a,\alpha}\hat{G}_{\alpha,a}\right)_N^{(B)}
  ,
\end{equation}
where the $A$- and $B$-type terms take the following form,
respectively:
\begin{widetext}
\begin{eqnarray}
 \label{eq:(A)-def}
 &&\left(\hat{\Sigma}_{a,\alpha}\hat{G}_{\alpha,a}\right)_N^{(A),1,1}
 =\\\nonumber
&& \hat{\Sigma}_{a,\alpha}^{1,1} \hat{g}_{\alpha,\gamma}^{1,1}
 \hat{\Sigma}_{\gamma,c}^{1,1} \hat{g}_{c,c'}^{1,1}
 \hat{\Sigma}_{c',\gamma'}^{1,1} \hat{g}_{\gamma',\delta'}^{1,1}
 \hat{\Sigma}_{\delta',d'}^{1,1} \hat{g}_{d',d}^{1,2}
 \hat{\Sigma}_{d,\delta}^{2,2} \hat{g}_{\delta,\beta}^{2,2}
 \hat{\Sigma}_{\beta,b}^{2,2} \hat{g}_{b,b}^{2,1}
 \hat{\Sigma}_{b,\beta}^{1,1} \hat{g}_{\beta,\delta}^{1,1}
 \hat{\Sigma}_{\delta,d}^{1,1} \hat{g}_{d,d'}^{1,1}
 \hat{\Sigma}_{d',\delta'}^{1,1} \hat{g}_{\delta',\gamma'}^{1,1}
 \hat{\Sigma}_{\gamma',c'}^{1,1} \hat{g}_{c',c}^{1,2}
 \hat{\Sigma}_{c,\gamma}^{2,2} \hat{g}_{\gamma,\alpha}^{2,2}
 \hat{\Sigma}_{\alpha,a}^{2,2} \hat{g}_{a,a}^{2,1}\\
 \nonumber
&+&
\hat{\Sigma}_{a,\alpha}^{1,1} \hat{g}_{\alpha,\delta}^{1,1}
 \hat{\Sigma}_{\delta,d}^{1,1} \hat{g}_{d,d'}^{1,1}
 \hat{\Sigma}_{d',\delta'}^{1,1} \hat{g}_{\delta',\gamma'}^{1,1}
 \hat{\Sigma}_{\gamma',c'}^{1,1} \hat{g}_{c',c}^{1,2}
 \hat{\Sigma}_{c,\gamma}^{2,2} \hat{g}_{\gamma,\beta}^{2,2}
 \hat{\Sigma}_{\beta,b}^{2,2} \hat{g}_{b,b}^{2,1}
 \hat{\Sigma}_{b,\beta}^{1,1} \hat{g}_{\beta,\gamma}^{1,1}
 \hat{\Sigma}_{\gamma,c}^{1,1} \hat{g}_{c,c'}^{1,1}
 \hat{\Sigma}_{c',\gamma'}^{1,1} \hat{g}_{\gamma',\delta'}^{1,1}
 \hat{\Sigma}_{\delta',d'}^{1,1} \hat{g}_{d',d}^{1,2}
 \hat{\Sigma}_{d,\delta}^{2,2} \hat{g}_{\delta,\alpha}^{2,2}
 \hat{\Sigma}_{\alpha,a}^{2,2} \hat{g}_{a,a}^{2,1}\\
 \label{eq:(B)-def}
 &&  \left(\hat{\Sigma}_{a,\alpha}\hat{G}_{\alpha,a}\right)_N^{(B),1,1}=\\
 \nonumber
&& \hat{\Sigma}_{a,\alpha}^{1,1} \hat{g}_{\alpha,\gamma}^{1,1}
 \hat{\Sigma}_{\gamma,c}^{1,1} \hat{g}_{c,c'}^{1,2}
 \hat{\Sigma}_{c',\gamma'}^{2,2} \hat{g}_{\gamma',\delta'}^{2,2}
 \hat{\Sigma}_{\delta',d'}^{2,2} \hat{g}_{d',d}^{2,2}
 \hat{\Sigma}_{d,\delta}^{2,2} \hat{g}_{\delta,\beta}^{2,2}
 \hat{\Sigma}_{\beta,b}^{2,2} \hat{g}_{b,b}^{2,1}
 \hat{\Sigma}_{b,\beta}^{1,1} \hat{g}_{\beta,\delta}^{1,1}
 \hat{\Sigma}_{\delta,d}^{1,1} \hat{g}_{d,d'}^{1,2}
 \hat{\Sigma}_{d',\delta'}^{2,2} \hat{g}_{\delta',\gamma'}^{2,2}
 \hat{\Sigma}_{\gamma',c'}^{2,2} \hat{g}_{c',c}^{2,2}
 \hat{\Sigma}_{c,\gamma}^{2,2} \hat{g}_{\gamma,\alpha}^{2,2}
 \hat{\Sigma}_{\alpha,a}^{2,2} \hat{g}_{a,a}^{2,1}\\
 \nonumber
 &+&
 \hat{\Sigma}_{a,\alpha}^{1,1} \hat{g}_{\alpha,\delta}^{1,1}
 \hat{\Sigma}_{\delta,d}^{1,1} \hat{g}_{d,d'}^{1,2}
 \hat{\Sigma}_{d',\delta'}^{2,2} \hat{g}_{\delta',\gamma'}^{2,2}
 \hat{\Sigma}_{\gamma',c'}^{2,2} \hat{g}_{c',c}^{2,2}
 \hat{\Sigma}_{c,\gamma}^{2,2} \hat{g}_{\gamma,\beta}^{2,2}
 \hat{\Sigma}_{\beta,b}^{2,2} \hat{g}_{b,b}^{2,1}
 \hat{\Sigma}_{b,\beta}^{1,1} \hat{g}_{\beta,\gamma}^{1,1}
 \hat{\Sigma}_{\gamma,c}^{1,1} \hat{g}_{c,c'}^{1,2}
 \hat{\Sigma}_{c',\gamma'}^{2,2} \hat{g}_{\gamma',\delta'}^{2,2}
 \hat{\Sigma}_{\delta',d'}^{2,2} \hat{g}_{d',d}^{2,2}
 \hat{\Sigma}_{d,\delta}^{2,2} \hat{g}_{\delta,\alpha}^{2,2}
 \hat{\Sigma}_{\alpha,a}^{2,2} \hat{g}_{a,a}^{2,1}
 .
\end{eqnarray}
Evaluating those terms, and the corresponding ``2,2'' Nambu
component, leads to the following:
\begin{eqnarray}
 \nonumber
 &&
 \overline{\left(\hat{\Sigma}_{a,\alpha}\hat{G}_{\alpha,a}\right)_N^{(A),1,1,+,-}(\omega)}
 -
  \overline{\left(\hat{\Sigma}_{a,\alpha}\hat{G}_{\alpha,a}\right)_N^{(A),2,2,+,-}(\omega)}
 =\\ \nonumber
&& i\frac{\Sigma_a^2 \Sigma_b^2 \Sigma_c^2 \Sigma_{c'}^2 \Sigma_d^2 \Sigma_{d'}^2}
    {2^4 W^{12}}\sin\left(-\varphi_a-\varphi_b+\varphi_c+\varphi_d\right)
    \left[\frac{1}{(k_F R_{\alpha,\gamma})(k_F R_{\gamma',\delta'})(k_F R_{\beta,\delta})}
 \cos\left(\frac{2\omega R_{\alpha,\gamma}}{\hbar v_F}\right)
 \cos\left(\frac{2\omega R_{\beta,\delta}}{\hbar v_F}\right)+\right.\\ \nonumber
 &&\left.
 \frac{1}{(k_F R_{\alpha,\delta})(k_F R_{\gamma',\delta'})(k_F R_{\beta,\gamma})}
 \cos\left(\frac{2\omega R_{\alpha,\delta}}{\hbar v_F}\right)
 \cos\left(\frac{2\omega R_{\beta,\gamma}}{\hbar v_F}\right)\right]\times\\ \nonumber
&& \left[\left(Y^A\right)^6 \left(\frac{\omega-i\eta}{\Delta}\right)^2
  -2 \left(Y^A\right)^5 Y^R \left|\frac{\omega-i\eta}{\Delta}\right|^2
  +2 \left(Y^A\right)^4 \left(Y^R\right)^2 \left|\frac{\omega-i\eta}{\Delta}\right|^2
  -2 \left(Y^A\right)^3 \left(Y^R\right)^3 \left|\frac{\omega-i\eta}{\Delta}\right|^2
  \right.\\
  &&\left.
  +2 \left(Y^A\right)^2 \left(Y^R\right)^4 \left(\frac{\omega+i\eta}{\Delta}\right)^2
  -2 Y^A \left(Y^R\right)^5 \left(\frac{\omega+i\eta}{\Delta}\right)^2
  +\left(Y^R\right)^6 \left(\frac{\omega+i\eta}{\Delta}\right)^2\right]
    \label{eq:terme-A}
 .
\end{eqnarray}
We also find
\begin{eqnarray} \nonumber
&&
 \overline{\left(\hat{\Sigma}_{a,\alpha}\hat{G}_{\alpha,a}\right)_N^{(B),1,1}(\omega)}
 -
 \overline{\left(\hat{\Sigma}_{a,\alpha}\hat{G}_{\alpha,a}\right)_N^{(B),2,2}(\omega)}\\
 \nonumber
 &=&
 i\frac{\Sigma_a^2 \Sigma_b^2 \Sigma_c^2 \Sigma_{c'}^2 \Sigma_d^2 \Sigma_{d'}^2}
    {2^4 W^{12}}\sin{\left(-\varphi_a-\varphi_b+\varphi_c+\varphi_d\right)}
    \left[\frac{1}{(k_F R_{\alpha,\gamma})(k_F R_{\gamma',\delta'})(k_F R_{\beta,\delta})}
 \cos\left(\frac{2\omega R_{\alpha,\gamma}}{\hbar v_F}\right)
 \cos\left(\frac{2\omega R_{\beta,\delta}}{\hbar v_F}\right)+\right.\\
\nonumber
 &&\left.
 \frac{1}{(k_F R_{\alpha,\delta})(k_F R_{\gamma',\delta'})(k_F R_{\beta,\gamma})}
 \cos\left(\frac{2\omega R_{\alpha,\delta}}{\hbar v_F}\right)
 \cos\left(\frac{2\omega R_{\beta,\gamma}}{\hbar v_F}\right)\right]\times\\\nonumber
&& \left[\left(Y^A\right)^6 \left(\frac{\omega-i\eta}{\Delta}\right)^2
  -2 \left(Y^A\right)^5 Y^R \left(\frac{\omega-i\eta}{\Delta}\right)^2
  +2 \left(Y^A\right)^4 \left(Y^R\right)^2 \left|\frac{\omega-i\eta}{\Delta}\right|^2
  -2 \left(Y^A\right)^3 \left(Y^R\right)^3 \left|\frac{\omega-i\eta}{\Delta}\right|^2
  \right.\\
  &&\left.
  +2 \left(Y^A\right)^2 \left(Y^R\right)^4 \left|\frac{\omega+i\eta}{\Delta}\right|^2
  -2 Y^A \left(Y^R\right)^5 \left(\frac{\omega+i\eta}{\Delta}\right)^2
  +\left(Y^R\right)^6 \left(\frac{\omega+i\eta}{\Delta}\right)^2\right]
    ,
    \label{eq:terme-B}
\end{eqnarray}
\end{widetext}
where the notations $Y^A$ and $Y^R$ are given by Eq.~(\ref{eq:YA}) and
Eq.~(\ref{eq:YR}), respectively. The spectral current $I_a(\omega)$ is
given by Eq.~(\ref{eq:Ia}):
\begin{eqnarray}
 I_a(\omega)&=&2\mbox{Re}\left[
  \overline{\left(\hat{\Sigma}_{a,\alpha}\hat{G}_{\alpha,a}\right)_N^{(A),1,1,+,-}(\omega)}
  \right.\\
  \nonumber
&& -\left.
 \overline{\left(\hat{\Sigma}_{a,\alpha}\hat{G}_{\alpha,a}\right)_N^{(A),2,2,+,-}(\omega)}
 \right]\\\nonumber
&& + 2 \mbox{Re} \left[
  \overline{\left(\hat{\Sigma}_{a,\alpha}\hat{G}_{\alpha,a}\right)_N^{(B),1,1,+,-}(\omega)}
  \right.\\
  \nonumber
&& -\left.
 \overline{\left(\hat{\Sigma}_{a,\alpha}\hat{G}_{\alpha,a}\right)_N^{(B),2,2,+,-}(\omega)}
 \right]
 ,
\end{eqnarray}
where each of those terms is given by
Eqs.~(\ref{eq:terme-A})-(\ref{eq:terme-B}).

To summarize, the split-quartet critical current takes the following
form at the lowest-order in the hopping amplitudes and in the
short-junction limit, i.e. if the conditions $\frac{2\Delta
R_{\alpha,\gamma}}{\hbar v_F} \alt 1$ and $\frac{2\Delta
R_{\beta,\delta}}{\hbar v_F} \alt 1$ are fulfilled:
\begin{equation}
 \label{eq:Isq-short}
 I_{sq}^{(short)}=I_{sq,4T,c}^{(short)}
 \sin\left(\varphi_a+\varphi_b-\varphi_c-\varphi_d\right) ,
\end{equation}
where, within the short-junction limit, the split-quartet critical
current $I_{sq,4T,c}^{(short)}$ is independent on the bias voltage
$V$. {Eq.~(\ref{eq:Isq-short}) is used in the main text, see
 Eq.~(\ref{eq:sq-short}).}

\subsection{Even-in-phase/odd-in-voltage four-terminal split-quartets}

\label{sec:sq-long}

We first make the Nambu structure explicit in the first term of
$\left(\hat{\Sigma}_{a,\alpha}\hat{G}_{\alpha,a}\right)_N^{(A),1,1}$,
see Eq.~(\ref{eq:(A)-def}):
\begin{eqnarray}
 \nonumber
&&\hat{\Sigma}_{a,\alpha}^{1,1} \hat{g}_{\alpha,\gamma}^{1,1}
 \hat{\Sigma}_{\gamma,c}^{1,1} \hat{g}_{c,c'}^{1,1}
 \hat{\Sigma}_{c',\gamma'}^{1,1} \hat{g}_{\gamma',\delta'}^{1,1}
 \hat{\Sigma}_{\delta',d'}^{1,1} \hat{g}_{d',d}^{1,2}
 \hat{\Sigma}_{d,\delta}^{2,2} \hat{g}_{\delta,\beta}^{2,2}
 \hat{\Sigma}_{\beta,b}^{2,2} \hat{g}_{b,b}^{2,1}\\
 &=& \Sigma_a^2 \Sigma_b^2 \Sigma_c^2 \Sigma_{c'}^2 \Sigma_{d'}^2
 g_{a,a}^{2,1} g_{b,b}^{2,1} \overline{g_{c,c'}^{1,1} g_{c',c}^{1,2}}\times
 \overline{g_{d,d'}^{1,1} g_{d',d}^{1,2}} \times Z
 \label{eq:(A)-non-eq}
 ,
\end{eqnarray}
with
\begin{eqnarray}
 \label{eq:(1)}
 Z&=&\overline{g_{\alpha,\gamma}^{1,1,+,-} g_{\gamma,\alpha}^{2,2,A}}
 \times \overline{g_{\gamma',\delta'}^{2,2,A} g_{\delta',\gamma'}^{2,2,A}}
 \times \overline{g_{\delta,\beta}^{2,2,A} g_{\beta,\delta}^{1,1,A}}\\
 \label{eq:(2)}
 &+&\overline{g_{\alpha,\gamma}^{1,1,R} g_{\gamma,\alpha}^{2,2,A}}
 \times \overline{g_{\gamma',\delta'}^{2,2,+,-} g_{\delta',\gamma'}^{2,2,A}}
 \times \overline{g_{\delta,\beta}^{2,2,A} g_{\beta,\delta}^{1,1,A}}\\
 \label{eq:(3)}
 &+&\overline{g_{\alpha,\gamma}^{1,1,R} g_{\gamma,\alpha}^{2,2,A}}
 \times \overline{g_{\gamma',\delta'}^{2,2,R} g_{\delta',\gamma'}^{2,2,A}}
 \times \overline{g_{\delta,\beta}^{2,2,+,-} g_{\beta,\delta}^{1,1,A}}\\
 \label{eq:(4)}
  &+&\overline{g_{\alpha,\gamma}^{1,1,R} g_{\gamma,\alpha}^{2,2,A}}
 \times \overline{g_{\gamma',\delta'}^{2,2,R} g_{\delta',\gamma'}^{2,2,A}}
 \times \overline{g_{\delta,\beta}^{2,2,R} g_{\beta,\delta}^{1,1,+,-}}\\
 \label{eq:(5)}
 &+&\overline{g_{\alpha,\gamma}^{1,1,R} g_{\gamma,\alpha}^{2,2,A}}
 \times \overline{g_{\gamma',\delta'}^{2,2,R} g_{\delta',\gamma'}^{2,2,+,-}}
 \times \overline{g_{\delta,\beta}^{2,2,R} g_{\beta,\delta}^{1,1,A}}\\
 \label{eq:(6)}
 &+&\overline{g_{\alpha,\gamma}^{1,1,R} g_{\gamma,\alpha}^{2,2,+,-}}
 \times \overline{g_{\gamma',\delta'}^{2,2,R} g_{\delta',\gamma'}^{2,2,R}}
 \times \overline{g_{\delta,\beta}^{2,2,R} g_{\beta,\delta}^{1,1,A}}
 .
\end{eqnarray}
Injecting the expression of the bare advanced and retarded Green's
functions of the ballistic 2D-metal into
Eqs.~(\ref{eq:(1)})-(\ref{eq:(6)}), see
Eqs.~(\ref{eq:Green-bare-debut})-(\ref{eq:Green-bare-fin}), leads to
(\ref{eq:(1)})$+$(\ref{eq:(4)})$=0$, (\ref{eq:(2)})$+$(\ref{eq:(3)})$=0$
and (\ref{eq:(5)})$+$(\ref{eq:(6)})$=0$, leading to $Z=0$ in
Eqs.~(\ref{eq:(1)})$-$(\ref{eq:(6)}) and to the full cancellation
(\ref{eq:(A)-non-eq})$=0$. The second term in Eq.~(\ref{eq:(A)-def})
is treated similarly, and we deduce $\left(\hat{\Sigma}_{a,\alpha}
\hat{G}_{\alpha,a}\right)_N^{(A),1,1} = 0$.

Concerning the first term appearing in
$\left(\hat{\Sigma}_{a,\alpha}\hat{G}_{\alpha,a}\right)_N^{(B),1,1}$,
see Eq.~(\ref{eq:(B)-def}), we obtain
\begin{eqnarray}
 \nonumber
&&\hat{\Sigma}_{a,\alpha}^{1,1} \hat{g}_{\alpha,\delta}^{1,1}
 \hat{\Sigma}_{\delta,d}^{1,1} \hat{g}_{d,d'}^{1,1}
 \hat{\Sigma}_{d',\delta'}^{1,1} \hat{g}_{\delta',\gamma'}^{1,1}
 \hat{\Sigma}_{\gamma',c'}^{1,1} \hat{g}_{c',c}^{1,2}
 \hat{\Sigma}_{c,\gamma}^{2,2} \hat{g}_{\gamma,\beta}^{2,2}
 \hat{\Sigma}_{\beta,b}^{2,2} \hat{g}_{b,b}^{2,1}\\
 &=& \Sigma_a^2
 \Sigma_b^2 \Sigma_c^2 \Sigma_{c'}^2 \Sigma_{d'}^2 g_{a,a}^{2,1}
 g_{b,b}^{2,1} \overline{g_{c,c'}^{1,1} g_{c',c}^{1,2}}\times
 \overline{g_{d,d'}^{1,1} g_{d',d}^{1,2}} \times Z 
 \label{eq:(B)-non-eq}
 .
\end{eqnarray}
We already demonstrated that $Z=0$, which leads to
(\ref{eq:(B)-def})$=0$.

It is concluded that (\ref{eq:Sigma-G})$=0$ leads to the vanishingly
small even-in-phase/odd-in-voltage four-terminal split-quartet
supercurrent, within the considered lowest-order perturbation theory
in the hopping amplitudes.
 
\section{Averaging over the short-scale oscillations}
\label{app:averaging}

In this Appendix, we detail how products of two Green's functions
propagating through the 2D-central-$N$ are averaged out over the
oscillations at the short scale of the Fermi wave-length.

In the limit of a two-dimensional electronic continuum, the ``1,1''
and ``2,2'' components of the bare Green's functions take the
following form as a function of the separation $R_0$ in real space:
\begin{eqnarray}
 \label{eq:Green-bare-debut}
 \hat{g}^{1,1,A,R}(R_0)&=&\frac{\pm i}{W} J_0\left(k_e R_0\right)\\
 \hat{g}^{2,2,A,R}(R_0)&=&\frac{\pm i}{W} J_0\left(k_h R_0\right)
 ,
\end{eqnarray}
with $J_0$ a Bessel function, and with $k_{e,h}=k_F \pm \frac{\delta
\mu_N}{\hbar v_F}$ the wave-vectors of the electron- and hole-type
excitations, where $\delta \mu_N$ is the shift in energy with respect
to the Fermi level. Assuming that the separation $R_0$ is much larger
than the energy-dependent Fermi wave-length, we obtain the following
approximation to the Nambu components of the one-particle bare
Green's functions:
\begin{eqnarray}
 \hat{g}^{1,1,A,R}(R_0)&\simeq&\frac{\pm i}{W\sqrt{k_F R_0}} \cos\left(k_e
 R_0 - \frac{\pi}{4}\right)\\ \hat{g}^{2,2,A,R}(R_0)&\simeq&\frac{\pm
  i}{W\sqrt{k_F R_0}} \cos\left(k_h R_0 - \frac{\pi}{4}\right)
 .
 \label{eq:Green-bare-fin}
\end{eqnarray}

Evaluating expressions like Eqs.~(\ref{eq:green1})-(\ref{eq:green4})
requires performing spatial averaging of the following type:
\begin{widetext}
\begin{equation}
 \overline{g^{1,1,A}(R_0) g^{2,2,A}(R_0)}
 \simeq-\frac{1}{W^2 (k_F R_0)}
 \overline{\cos\left(k_F R_0 - \frac{\pi}{4}+\frac{\delta \mu_N R_0}{\hbar v_F}\right)
  \cos\left(k_F R_0 - \frac{\pi}{4}-\frac{\delta \mu_N R_0}{\hbar v_F}\right)}
 .
\end{equation}
Expanding
\begin{equation}
 \cos\left(k_F R_0 - \frac{\pi}{4}\pm\frac{\delta \mu_N R_0}{\hbar v_F}\right)
 = \cos\left(k_F R_0 - \frac{\pi}{4}\right) \cos\left(\frac{\delta \mu_N R_0}{\hbar v_F}\right)
 \mp
 \sin\left(k_F R_0 - \frac{\pi}{4}\right) \sin\left(\frac{\delta \mu_N R_0}{\hbar v_F}\right)
\end{equation}
leads to the following expression for the average over the fast
variable $k_F R_0 - \frac{\pi}{4}$:
\begin{equation}
  \overline{g^{1,1,A}(R_0) g^{2,2,A}(R_0)}\simeq
 -\frac{1}{2W^2 (k_F R_0)} \cos\left(\frac{2\delta \mu_N R_0}{\hbar v_F}\right)
 .
\end{equation}
\end{widetext}

\end{document}